\documentstyle[11pt,preprint,aps,prb]{revtex}
\draft
\begin{document}
\title{AUGER RECOMBINATION
IN SEMICONDUCTOR QUANTUM WELLS}
\vspace{1cm}
\author{Anatoli S. Polkovnikov and Georgy G. Zegrya  \\
{\it Ioffe Physico-Technical Institute}\\
Russia, 194021, St.-Petersburg, Politechnicheskaya st. 26}
\maketitle
\begin{abstract}
The principal mechanisms of Auger recombination of nonequilibrium carriers
in semiconductor heterostructures with quantum wells are investigated.
It is shown for
the first time that there exist three fundamentally different Auger
recombination mechanisms
of (i) thresholdless, (ii) quasi-threshold, and (iii) threshold types.
The rate of the thresholdless Auger process depends on temperature
only slightly. The rate of
the quasi-threshold Auger process depends on temperature exponentially.
However, its threshold energy essentially varies with quantum well width
and is close to zero for narrow quantum wells. It is shown that the
thresholdless and the quasi-threshold Auger processes dominate in narrow
quantum wells, while the threshold and the quasi-threshold processes prevail
in wide quantum wells. The limiting case of a three-dimensional (3D)
Auger process is reached for infinitely wide quantum wells.
The critical quantum well width is
found at which the quasi-threshold and threshold Auger processes merge into
a single 3D Auger process.  Also studied is phonon-assisted Auger
recombination in quantum wells.
It is shown that for narrow quantum wells the act of phonon emission becomes
resonant, which in turn increases substantially the coefficient of
phonon-assisted Auger recombination. Conditions are found under which the
direct Auger process dominates over the phonon-assisted Auger
recombination at various temperatures and
quantum well widths.
\end{abstract}
{\begin{center}
\section*{\normalsize {\bf I.~Introduction}}
\end{center}}

Two recombination processes are predominant in semiconductors at high
excitation levels: (i) radiative recombination and (ii) nonradiative Auger
recombination related to the electron-electron interaction. For homogeneous
semiconductors, mechanisms of Auger recombination (AR) have been extensively
studied~\cite{1}$^-$~\cite{3a}. In narrow gap semiconductors
occur AR processes
involving two electrons and a heavy hole (CHCC Auger process) or an electron
and two heavy holes, with transition of one of the holes to the SO zone
(CHHS Auger process)~\cite{2,3a,4}. Both these processes are of threshold
nature, and the rate of Auger recombination changes with temperature
exponentially ~\cite{1,2}. The only exception are semiconductors in which
the spin-orbit splitting is close to the energy gap (GaSb and InAs). Under
certain conditions the rate of the CHHS process in these semiconductors
depends on temperature only slightly~\cite{4a}.
It is commonly believed that in
weakly doped semiconductors phonon-assisted AR processes are predominant
at low temperature and high excitation levels~\cite{3,4}. Because of
the large momentum transferred to a phonon, the threshold for heavy holes
is removed and phonon-assisted Auger process becomes a power-law function
of temperature.

Single semiconductor heterostructures, quantum wells, quantum wires, quantum
dots are spatially inhomogeneous owing to the existence of barriers.
The presence of a heteroboundary affects not only the energy and wave
functions of carriers, but also the macroscopic properties of heterostructures
~\cite{5}, which is of primary importance. It is
commonly believed that the AR mechanism in quantum wells is the same as in a
homogeneous semiconductor ~\cite{4,6}$^-$\cite{9}. Nevertheless, the presence
of a heteroboundary strongly affects the electron-electron interaction in
quantum wells, and this influence is of fundamental nature. The
heteroboundary lifts restrictions imposed on processes of electron-electron
interaction by the energy and momentum conservation laws. Namely, the
conservation of quasi-momentum perpendicular to the heteroboundary breaks
down. In turn, this leads to the appearance in heterostructures of new
thresholdless channels of Auger recombination~\cite{5}. The rate of the
thresholdless AR process is a power function of temperature. 
The first direct experiment aimed at observing the thresholdless AR
channel at $T=77K$ was reported in~\cite{10}. At low
temperature the thresholdless process, in contrast to that with a threshold,
is rather an efficient channel of nonradiative recombination, and
for narrow quantum wells and high concentrations it dominates the
phonon-assisted AR process. The existence of a thresholdless matrix element
of electron-electron interaction also exerts strong influence on the
phonon-assisted AR process. The latter becomes resonant and is much
enhanced as compared with the 3D phonon-assisted Auger
process.

A detailed analysis of the threshold and thresholdless AR mechanisms has been
performed for a single heterobarrier~\cite{5}. Conditions were studied under
which the thresholdless channel dominates over the threshold one.
For quantum wells
no such detailed analysis has been done. Taylor {\it et al.}~\cite{10a}
considered the possibility of removing the threshold for AR in
quantum wells upon
transition of excited carriers to the continuos part of spectrum. However,
no microscopic theory of the thresholdless process was given in this work and
no theoretical analysis performed of the competition between the threshold,
quasi-threshold, and thresholdless AR mechanisms at various temperatures and
quantum well widths. Only the thresholdless AR channel, corresponding to small
momenta transferred in Coulomb interaction of particles (for the CHCC process)
with spin-orbit interaction neglected, was considered in~\cite{11,12}.

The aim of the present work is to investigate theoretically the principal
mechanisms of AR for nonequilibrium carriers in semiconductor quantum wells.
It will be shown
that there exist three fundamentally different AR mechanisms: (i) threshold
mechanism similar to an Auger process in a homogeneous semiconductor, (ii)
quasi-threshold mechanism with a threshold energy strongly depending on the
quantum well width, and (iii) thresholdless mechanism inoperative in a
homogeneous semiconductor. For the threshold AR process in a quantum well
the threshold energy is close to that in a homogeneous semiconductor.
Conversely, owing to the small value of the threshold energy, the rate of
the quasi-threshold process in narrow quantum wells depends on temperature
only slightly. For this reason there is no clear distinction between
mechanisms (ii) and (iii) in sufficiently narrow quantum wells, and they
may be considered as a single thresholdless AR process. With increasing
quantum well width, the threshold energy of the quasi-threshold process
increases and approaches the bulk value. A totally different behavior is
characteristic of the thresholdless AR mechanism. With increasing quantum
well width, its rate falls dramatically and, on passing to a homogeneous
semiconductor, this mechanism ceases to be
operative. Conditions will be found under which the thresholdless AR
mechanism dominates the threshold one. The critical quantum well width
will be found , at which the quasi-threshold and the threshold Auger
processes merge and form a single 3D AR process. Also, the phonon-assisted
AR in quantum wells will be studied. It will be shown that for narrow
quantum wells the act of phonon emission becomes resonant, which in turn
increases substantially the coefficient of phonon-assisted AR. Conditions
will be found under which the direct AR process dominates the phonon-assisted
AR process at various temperatures and quantum well widths.

\begin{center}\section*{\bf II.~Principal Equations}\end{center}

To analyze the AR mechanisms and find the rate of the Auger process, wave
functions of charge carriers are to be known. As already established for bulk
Auger processes, the wave functions of carriers must be calculated using the
multiband approximation~\cite{2}. We will use the four-band Kane's model,
the most adequately describing the wave functions and energy spectrum of
carriers in narrow-gap A$_{III}$B$_V$ semiconductors~\cite{13}.

\begin{center}\subsection*{\bf 1.~Wave functions in a homogeneous
semiconductor}\end{center}

For most A$_{III}$B$_V$ semiconductors, wave functions of electrons and holes
in the center of the Brillouin zone are described by the $\Gamma_6^+$
representation for the conduction band and by the $\Gamma_7^+$ and
$\Gamma_8^+$ representations for the valence band. Of these the first two
and the last are doubly and fourfold degenerate, respectively.
The corresponding
equations for wave functions may be written in differential form. Commonly,
the basis wave functions of the conduction and valence bands are taken in
form of eigenfunctions of the angular momentum~\cite{13,15}. However, another
representation of the basis functions is more appropriate for our purposes:
\begin{equation}
|s\uparrow\rangle\;, |s\downarrow\rangle\;,
|x\uparrow\rangle\;, |x\downarrow\rangle\;,
|y\uparrow\rangle\;, |y\downarrow\rangle\;,
|z\uparrow\rangle\;, |z\downarrow\rangle\;,
\label{eq:1}
\end{equation}
where $|s\rangle$ and $|x\rangle,\,|y\rangle,|z\rangle$ are the Bloch
functions of $s$ and ${\bf p}$ type with angular momenta of 0 and 1,
respectively. The
former describe the state of the conduction band and the latter the state of
the valence band at the $\Gamma$-point. Arrows denote the direction of 
spin. The wave function of carriers $\psi$ may be presented in the form:
\begin{displaymath}
\psi=\Psi_s|s\rangle+{\bbox \Psi}{|\bf p}\rangle,
\end{displaymath}
where $\Psi_s$ and  $\bbox \Psi$ are spinors.
In the vicinity of the $\Gamma$-point the equations for $\Psi_s$ and
${\bbox\Psi}$ envelopes written in the spherical approximation are as follows:
\begin{equation}
\left\{
\begin{array}{l}
(E_c-E)\Psi_s-i\hbar\gamma\nabla\!{\bf\Psi}=0,\\
(E_v-\delta-E){\bf\Psi}-i\hbar\gamma\nabla\!\Psi_s+\frac{\hbar^2}{2m}
(\tilde{\gamma}_1+4\tilde{\gamma}_2)\nabla\!(\nabla\!{\bf\Psi})-\\
-\frac{\hbar^2}{2m}(\tilde{\gamma}_1-2\tilde{\gamma}_2)\nabla\!\times\!
[\nabla\!\times\!{\bf\Psi}]+i\delta[{\bbox{\sigma}}\times{\bf\Psi}]=0.
\end{array}
\label{eq:2}
\right.
\end{equation}
Here $\gamma$ is the Kane's matrix element~\cite{15} having
dimension of velocity, $\widetilde{\gamma}_1$ and $\widetilde{\gamma}_2
=\widetilde{\gamma}_3$ are the generalized Luttinger parameters~\cite{15},
$\textstyle \delta=\Delta_{so}/ 3$, $\Delta_{so}$ is the spin orbit splitting,
$E_v$ and  $E_c$ are the energies of the lower edge of the conduction band and
the upper edge of the valence band, $m$ is the free-electron mass,
${\bbox{\sigma}}=(\sigma_x,\sigma_y,\sigma_z)$ are the Pauli spin matrices.
If, instead of using the Luttinger parameters, the heavy hole mass describing
the interaction with higher bands is introduced phenomenologically,
then equations (\ref{eq:2}) are transformed into equations derived by Suris
~\cite{14}. It is easy to verify that equations (\ref{eq:2}) are identical
to those commonly used~\cite{15,16}$^-$\cite{18}. In the first equation
in the system (\ref{eq:2}) we neglect the term with mass of free electrons.

\subsubsection*{\begin{center}\rm A.~ Hole states\end{center}}

The expression for ${\Psi}_s$ can be found from the first equation of the
system (\ref{eq:2}). Substitution of ${\Psi}_s$ into the second equation
gives:
\begin{equation}
-E\bbox{\Psi}+{\hbar^2\over
2m_l}\nabla(\nabla\!\bbox{\Psi})-{\hbar^2\over
2m_h}\nabla\!\times\![\nabla\!\times\!\bbox{\Psi}]+i\delta[\bbox{\sigma}
\times\bbox{\Psi}]=0,
\label{eq:3}
\end{equation}
where
\begin{displaymath}
m_l^{-1}=
{2\gamma^2\over E_g+\delta-E}
+m^{-1}
(\widetilde{\gamma}_1+4\widetilde{\gamma}_2),\quad
m_h^{-1}=m^{-1}
(\widetilde{\gamma}_1-2\widetilde{\gamma}_2).
\end{displaymath}
Here $m_h$ coincides with the heavy hole mass, and $m_l$ with the light hole
mass in the case of zero constant of spin-orbit interaction; $E_g=E_c-E_v$ is
the semiconductor forbidden gap. For the sake of convenience, it is assumed
that $E_v=\delta$. This choice is due to an increase in the heavy hole 
and light hole energies at the $\Gamma$-point by $\delta$
and a decrease in the \underline{spin split off} (SO) hole
energy by $2\delta$ under the action of spin-orbital interaction
(see eq. \ref{eq:7}). Equation (\ref{eq:3}) can be simplified by introducing
new functions
\begin{equation}
\phi=div\bbox{\Psi}\;\;\mbox{and }\;\;\eta=\bbox{\sigma}\,rot\bbox{\Psi}.
\label{eq:4}
\end{equation}

After taking the divergence and rotor of equation (\ref{eq:3}), multiplied
by ${\bf{\sigma}}$ it is transformed into a system of two differential
equations
\begin{equation}
\left\{
\begin{array}{l}
-E\phi+\frac{\hbar^2}{2m_l}\Delta\!\phi+i\delta\eta=0\; ,\\
-(E+\delta)\eta+\frac{\hbar^2}{2m_h}\Delta\!\eta-2i\delta\phi=0.
\end{array}
\right.
\label{eq:5}
\end{equation}

Fourier transform of these equations gives hole spectra for a homogeneous
semiconductor
\begin{equation}
\left[
\begin{array}{cc}
E+\frac{\hbar^2}{2m_l}k^2 & i\delta\\
-2i\delta & E+\frac{\hbar^2}{2m_h}k^2+\delta
\end{array}
\right]
\left(
\begin{array}{c}
\phi \\ \eta
\end{array}
\right)
=0.
\label{eq:6}
\end{equation}

The characteristic equation has two roots
\begin{equation}
E_{1,2}=-\frac{\delta}{2}-\frac{\hbar^2  k^2}{4}(m_l^{-1}+m_h^{-1})\pm
\sqrt{2\delta^2+\left(\frac{\delta}{2}-\frac{\hbar^2 k^2}{4}(m_l^{-1}
-m_h^{-1})\right)^2}.
\label{eq:7}
\end{equation}

It should be noted that $m_l$ depends on energy (see eq. (\ref{eq:3})). At the
$\Gamma$-point ($k=0$) we have the roots  $E_1=\delta$ and $E_2=-2\delta$.
The positive solution corresponds to light holes, and that
with negative sign to SO holes.

In the vicinity of the $\Gamma$-point the energies $E_{1,2}$ can be expanded
into a series in terms of wave vector to relate the effective light and
SO-hole masses $m_{hl},\;m_{so}$ and the Luttinger parameters:
\begin{eqnarray}
E_1\approx\delta-\frac{\hbar^2 k^2}{2m_{hl}},\;\;
E_2\approx-2\delta-\frac{\hbar^2 k^2}{2m_{so}},
\label{eq:9}
\end{eqnarray}
where
\begin{eqnarray*}
m_{hl}^{-1}=\frac{4\gamma^2}{3E_g}+\frac{(\tilde{\gamma}_1
+2\tilde{\gamma}_2)}{m},\;\;
m_{so}^{-1}=\frac{2\gamma^2}{3(E_g+3\delta)}+\frac{\tilde{\gamma}_1}{m}.
\end{eqnarray*}
An approximate light hole spectrum can be obtained by means of a widely used
$4\times 4$ Hamiltonian~\cite{17}. However, the range of its applicability is
rather narrow, since commonly $m_l\sim 0.1\,m_h$ and the expansion
(\ref{eq:9}) is only valid when $E\ll {m_l\over m_h}\Delta_{so}$. Moreover,
such a model cannot describe Auger transitions at all, since the basis
states of
carriers in different bands are orthogonal. The same applies to the spectrum
of the SO band.

The Fourier amplitudes of the wave functions of both light and SO
holes can be presented in the form (see eq. (\ref{eq:3})):
\begin{eqnarray}
\bbox{\Psi}={\bf k}f+\frac{i\delta}{E+\delta+\frac{\hbar^2k^2(E)}{2m_h}}
\left[{\bf k}\times\bbox{\sigma}f\right],\;\;
\Psi_s=-\frac{\hbar\gamma k^2(E)}{E_g+\delta-E}f,
\label{eq:10}
\end{eqnarray}
where $f$ is an arbitrary spinor related to the previously introduced
function $\phi$ by the expression $\phi =k^2(E)f$.

The third solution of (\ref{eq:3}) pertaining to heavy holes satisfies the
relations $div{\bf\Psi}=0$ (as a consequence $\Psi_s=0$)
and ${\bbox{\sigma}}{\rm rot}{\bf\Psi}=0$.
This follows from equation (\ref{eq:5}), since, if $\phi=0$,
then $\eta=0$ and vice versa. It can be readily seen that
\begin{displaymath}
\left[\bbox{\sigma}\times\bbox{\Psi}_h\right]=-i\bbox{\Psi}_h.
\end{displaymath}

Thus, the dispersion law describing the heavy hole spectrum looks like
\begin{equation}
E_h=\delta-\frac{\hbar^2k_h^2}{2m_h}.
\label{eq:11}
\end{equation}

The components of the heavy hole
wave function must satisfy the equations:
\begin{eqnarray}
&&\left\{
\begin{array}{l}
\Psi_{z\downarrow}=(\Psi_{x\uparrow}+i\Psi_{y\uparrow})\;\\
\Psi_{z\uparrow}=(-\Psi_{x\downarrow}+i\Psi_{y\downarrow})
\end{array}
\right.
\;\Leftrightarrow\;[\bbox{\sigma}\times\bbox{\Psi}]=-i\bbox{\Psi},\\
&&\left\{
\begin{array}{l}
k_z\Psi_{z\uparrow}+k_x\Psi_{x\uparrow}+k_y\Psi_{y\uparrow}=0\\
k_z\Psi_{z\downarrow}+k_x\Psi_{x\downarrow}+k_y\Psi_{y\downarrow}=0
\end{array}
\right.
\;\Leftrightarrow\;div\bbox{\Psi}=0.
\label{eq:12}
\end{eqnarray}
Solving these equations one may obtain the explicit expressions
for the wave functions. For a quantum well they are given in Appendix A.

\subsubsection*{\begin{center}\rm B.~Electron states\end{center}}

In principle, the conventional equations for electrons have the same form as
those for holes. Since the $\Gamma$-point in the conduction band is only
doubly degenerate, and the crystal field causes no additional splitting,
there is no need to retain terms with parameters $\widetilde{\gamma}_i$.
Moreover, the presence of these terms in the equations for electrons
gives a far too exact model. Thus, a simplified model will be used for
electrons

\begin{equation}
\left\{ \begin{array}{l}
(E_c-E)\Psi_s-i\hbar\gamma\nabla\!\bbox{\Psi}=0,\\
(E_v-\delta-E)\bbox{\Psi}-i\hbar\gamma\nabla\!\Psi_s
+i\delta[\bbox{\sigma}\times
\bbox{\Psi}]=0.
\end{array}
\right.
\label{eq:16}
\end{equation}

The electron energies can be conveniently reckoned from the lower edge of the
conduction band ($E_c=0$). This energy will be denoted by ${\cal E}$, so that
it would not be confused with the full energy of electron $E$, reckoned from
the same level as the hole energy. Introducing into equation (\ref{eq:16})
the functions $\phi$ and $\eta$ in the same form as before
(see eq. (\ref{eq:4})), one can obtain
\begin{equation}
\begin{array}{l}
-(E_g+\delta+{\cal E})\phi+\frac{\hbar^2\gamma^2}{{\cal
E}}\Delta\phi+i\delta\eta=0,
\\-(E_g+{\cal E}+2\delta)\eta-2i\delta\phi=0.
\end{array}
\label{eq:17}
\end{equation}
Passing to a Fourier transform, we find the electron dispersion law
\begin{equation}
k^2=\frac{{\cal E}}{\hbar^2\gamma^2}\frac{{\cal E}^2+{\cal E}(2E_g+3\delta)+
(E_g+3\delta)E_g}{E_g+{\cal E}+2\delta}.
\label{eq:18}
\end{equation}
If ${\cal E}\ll E_g$, $\delta$, then the energy is quadratic in wave vector.
\begin{equation}
{\cal E}={\hbar^2k^2\over 2m_c},
\label{eq:19}
\end{equation}
where
\begin{displaymath}
m_c^{-1}=2\gamma^2\frac{E_g+2\delta}{(E_g+3\delta)E_g}.
\end{displaymath}
The Fourier amplitude of the wave function is given by
\begin{eqnarray}
\Psi_s=f,\;\;\bbox{\Psi}
=\frac{{\cal E}}{\hbar\gamma k^2({\cal E})}\left[{\bf k}f+
\frac{i\delta}{{\cal E}+E_g+2\delta}\left[{\bf k}\times\left(\bbox{\sigma}f
\right)\right]\right]\;,
\label{eq:20}
\end{eqnarray}
where $f$ is an arbitrary spinor (see eq. (\ref{eq:10})).

\subsubsection*{\begin{center}\normalsize\rm C.~Probability flux and
the equations near the heteroboundary\end{center}}

An expression for the probability flux density can be derived from equation
(\ref{eq:2}) by substituting $E\rightarrow -i\hbar{\partial\over\partial t}$
and using then a procedure similar to that employed in
quantum mechanics~\cite{19}. It can also be derived by the ${\bf kp}$ method
in the second-order perturbation theory. As a result, the following
expression is obtained in the case of holes for the probability flux density
\begin{equation}
{\bf j}_{h}=\frac{E_g+\delta-E}{2m_l\gamma}
[\Psi_s\bbox{\Psi}^*+\Psi_s^*\bbox{\Psi}]-\frac{i\hbar}{2m_h}
(\bbox{\Psi}\times{\rm
rot}\bbox{\Psi}^*-\bbox{\Psi}^*\times{\rm rot}\bbox{\Psi}).  \label{eq:22}
\end{equation}
For electrons in the conduction band this expression takes a simpler form
\begin{equation} {\bf j}_{e}=\gamma
[\Psi_s\bbox{\Psi}^*+\Psi_s^*\bbox{\Psi}].  \label{eq:23}
\end{equation}
The exact procedure for deriving boundary conditions for wave functions at
the heteroboundary still remains to be devised. However, some approximate
methods for solving this problem have been developed in recent years.
The Kane's parameter $\gamma$ usually differs only slightly for
$A_{III}B_V$ semiconductors, hence continuity of $\gamma$ is usually
supposed in literature (see for example~\cite{15}). Discrepancy
of parameter $\gamma$ in a quantum well and barrier region
results in small change of Auger coefficient (see section IV).
Following the method elaborated by Burt~\cite{16} and asuming
contunuity of Kane's parameter, we derive from the
system (\ref{eq:2}) Kane's equations which can be integrated across the
heterobarrier:
\begin{equation}
\left\{
\begin{array}{l}
(E_g+\delta-E)\Psi_s-i\hbar\gamma\nabla\!\bbox{\Psi}=0,\\
-E\bbox{\Psi}-i\hbar\gamma\nabla\!\Psi_s+\frac{\hbar^2}{2m}\nabla\!\left[
6\tilde{\gamma}_2\nabla\!\bbox{\Psi}\right]+\\
+{\hbar^2\over 2m}\,
{\partial\over\partial x_k}\left(\tilde{\gamma}_1-2\tilde{\gamma}_2\right)
\frac{\partial}{\partial x_k}\bbox{\Psi}+i\delta[\bbox{\sigma}\times
\bbox{\Psi}]=0.
\end{array}
\right.
\label{eq:24}
\end{equation}
Using these equations and the probability flux density conservation law we
derive the boundary conditions for the wave-function envelopes.

\begin{center}
\subsection*{\bf 2.~Carrier states in a quantum well}
\end{center}

The wave functions of carriers in a quantum well may be derived using the
symmetry properties of the Hamiltonian. Spinless Hamiltonian
${\cal H}_0$ is invariant with respect to the substitution
$x\rightarrow -x$. Consider an operator ${\cal R}$ such that
\begin{equation}
{\cal R}:(x,y,z)\rightarrow (-x,y,z),\quad {\cal R}={\cal IC}_{\pi x},
\label{eq:25}
\end{equation}
\begin{displaymath}
{\cal H}_0{\cal R}={\cal RH}_0,
\end{displaymath}
where ${\cal I}$ is the inversion operator, and ${\cal C}_{\pi x}$ is the
operator of rotation by an angle $\pi$ around the $x$ axis perpendicular
to the plane of the quantum well.

With account of the spin orbit interaction the Hamiltonian can be written in
the form:
\begin{equation}
{\cal H}={\cal H}_0+\frac{\hbar}{4m^2c^2}{[{\nabla V}
\times {\bf p}]}{\bbox{\sigma}},
\label{eq:26}
\end{equation}
where $\bf p$ is the momentum operator and $V$ is the potential
energy of an electron in the crystal. The last term does not commute with R.
Therefore, the symmetry operator ${\cal D}$ may be sought for as a product of
operator ${\cal R}$ and some spin matrix $S$ to be found:
${\cal D}={\cal R}\otimes S$. Since inversion leaves unchanged the sign
of the vector product, the matrix $S$ must satisfy the relations
\begin{equation}
\left\{
\begin{array}{c}
S\sigma_x=\sigma_xS \; \\
S\sigma_y=-\sigma_yS \;\\
S\sigma_z=-\sigma_zS  \;
\end{array}\;
\right. ,
\sigma_x=\left[ \begin{array}{cc} 0 & 1 \vspace{-2 mm}\\ 1 & 0\end{array}
\right]\; ,
\sigma_y=\left[ \begin{array}{cc} 0 & -i \vspace{-2 mm}\\ i & 0\end{array}
\right] \; ,
\sigma_z=\left[ \begin{array}{cc} 1 & 0 \vspace{-2 mm}\\ 0 & -1\end{array}
\right]\; .
\label{eq:27}
\end{equation}
Obviously a Pauli spin matrix $\sigma_x$ may be taken for the matrix $S$:
$S=\sigma_x$.

The functions $\Psi (x,y,z)$ and ${\cal D}\Psi (-x,y,z)$ satisfy the same
equation. For this reason the eigenfunctions of the Hamiltonian may be sought
for as eigenfunctions of the operator ${\cal D}$.
\begin{equation}
\Psi(x,y,z)=\nu\,{\cal D} \Psi(-x,y,z),\mbox{ where}\;\nu=\pm 1
\label{eq:30}
\end{equation}
The values $\nu=\pm 1$ correspond to carrier states with different symmetry.
With the wave functions chosen in such a way, the boundary conditions may be
satisfied at one heteroboundary only, since at the other they will be
fulfilled automatically. Solving equation (\ref{eq:30}) we find the
necessary conditions for various components of the symmetrized wave function.
\begin{eqnarray}
\Psi_{s\uparrow}(x,y,z)=\pm\Psi_{s\downarrow}(-x,y,z) \;\; ,
\Psi_{x\uparrow}(x,y,z)=\mp\Psi_{x\downarrow}(-x,y,z) \; ,\nonumber \\
\Psi_{y\uparrow}(x,y,z)=\pm\Psi_{y\downarrow}(-x,y,z) \;\; ,
\Psi_{z\uparrow}(x,y,z)=\pm\Psi_{z\downarrow}(-x,y,z) \;
\label{eq:31}\\
\mbox{where the sign $"+"$ corresponds to $\nu=1$, and  $"-"$ to $\nu=-1$}
\;\nonumber
\end{eqnarray}
Corresponding expressions for the components of
electron and hole wave functions are given in Appendix A.

\begin{center}
\section*{{\bf  III.~Probability of Auger recombination}}
\end{center}
The probability of AR per unit time can be calculated in terms of the
first-order perturbation theory in electron-electron interaction:
\begin{equation}
W_{i\rightarrow f}=
{2\pi\over\hbar}
|M_{fi}|^2
\delta (\varepsilon_f-\varepsilon_i),
\label{eq:50}
\end{equation}
where
\begin{equation}
M_{fi}=
\langle\Psi_f(\bbox{r}_1,\bbox{r}_2,\nu_1,\nu_2)
\left| {e^2\over\kappa_0|\bbox{r}_1-\bbox{r}_2|}
+\tilde{\Phi}({\bf r}_1,{\bf r}_2)\right|
\Psi_i(\bbox{r}_1,\bbox{r}_2,\nu_1,\nu_2)\,
\rangle
\label{51}
\end{equation}
is the matrix element of electron-electron interaction, $\bbox{r}_1$ and
$\bbox{r}_2$ are carrier coordinates, $\nu_1$ and $\nu_2$ are spin variables
(see eq. (\ref{eq:30})), $e$ is an electron charge, and $\kappa_0$ is the
dielectric constant of the intrinsic semiconductor,
$\tilde\Phi({\bf r}_1,{\bf r}_2)$ is the additional potential arising
because of difference between quantum well and barrier dielectric
constants. The explicit expressions for $\tilde\Phi({\bf r}_1,{\bf r}_2)$
are given in the Appendix B.

Taking into account the antisymmetrized form of the wave
functions, the matrix element of Auger transition is the following:
\begin{equation}
M_{fi}=M_{\rm I}-M_{\rm II},
\label{eq:53}
\end{equation}
where
\begin{equation}
\begin{array}{ll}
M_{\rm I}=
\langle\Psi_3(\bbox{r}_1,\nu_1)
\Psi_4(\bbox{r}_2,\nu_2)
\left|
{\textstyle e^2\over\textstyle\kappa_0|\bbox{r}_1-\bbox{r}_2|}
+\tilde\Phi({\bf r}_1,{\bf r}_2)
\right|
\Psi_1(\bbox{r}_1,\nu_1)
\Psi_2(\bbox{r}_2,\nu_2)
\rangle,
\end{array}
\label{eq:54}
\end{equation}
the expression for $M_{II}$ may be derived from (\ref{eq:54}) by
swaping indexes $1$ and $2$ in the wave functions $\Psi_1$ and $\Psi_2$.
Hereafter the indexes $I$ and $II$ in the expressions for the matrix
elements will be omitted.

We shall consider two AR processes, CHCC and CHHS, since in fact only these
two determine the rate of Auger recombination. Strictly speaking such a
terminology is inapplicable to carriers in a quantum well, since there
exists mixing between heavy hole, light hole and SO hole subbands.
However, as noted above, in the case $m_c\ll m_h$ the extent of mixing
between heavy and light holes is low, and the mixing of SO holes
with heavy and
light holes is negligible at $\Delta_{so}\gg T$. The last condition is
nearly always fulfilled for A$_{III}$B$_V$ semiconductors. For this reason
we may rely on the above terminology.

\subsection*{\begin{center} Matrix element of Auger transition
\end{center}}
Evaluations of matrix elements for the CHCC and CHHS Auger processes
are similar. For the sake of simplicity later in this section the
matrix element of the CHCC Auger transition will be mainly discussed.
However, in the following
section approximate expressions for the Auger coefficient will be given
both for the CHCC and for the CHHS processes.
The matrix element of an electron-electron Coulomb interaction can be most
conveniently calculated using a Fourier transform. We take into account
that the wave functions of carriers in a quantum well are the plane waves
along the lateral direction:
\begin{displaymath}
\Psi_i({\bf r})=\psi_i(x,{\bf q}_i)e^{i{\bf q}_i{\bbox\rho}}.
\end{displaymath}
The explicit expressions for wave functions of electrons and holes
$\psi_i$ are given in the Appendix A.
Then
\begin{eqnarray}
M&=&
{\textstyle 4\pi e^2\over\textstyle\kappa_0}{\textstyle 1\over
\textstyle 2q}
\int_{-\infty}^\infty\int_{-a/2}^{a/2}
\psi_4^\ast(x_1)\psi_3^\ast(x_2)\times\nonumber\\
&\times&\left(e^{-q|x_1-x_2|}
+\tilde\phi(x_1,x_2,q)\right)\psi_1(x_1)\psi_2(x_2)\;dx_1dx_2,
\label{eq:55}
\end{eqnarray}
$q=|q_1-q_4|=|q_3-q_2|$  is the momentum transferred in the plane of the
quantum well in Coulomb interaction, $a$ is the quantum well width
$\tilde\phi$ corresponds to the potential $\tilde\Phi$ and
the expression for it is given in the Appendix B. The integrating
over $x_2$ is limited within the quantum well due to the fact
that heavy holes, because of their relatively big mass, are usually
strongly localized inside the well.
Hereafter $x$ denotes the coordinate
orthogonal to the quantum well plain and ${\bbox \rho}$ denotes the
pair of coordinates in the quantum well plane, $\bf q$ and $k$
are the lateral and $x$- quasimomentum components of particles.

As it is seen from the equation (\ref{eq:55}) the Auger scattering
occurs on the one-dimensional exponentially decaying potential
which depends on the lateral transferred momentum.
The state of the excited particle may lie in both the
continuous and discrete spectrum, \footnote[1]{}
with the latter situation occurring when the
longitudinal momentum of the particle much exceeds the transverse momentum.
In determining the rate of Auger recombination, both localized and
delocalized states must be considered as final states of the
excited particle. Possibility of an electron (hole) transition into a
localized or into a free state leads to the existence of different
AR mechanisms in quantum wells.

\begin{center}
{\it Calculation of the matrix element of Auger recombination for a
transition of the excited particle into the continuous spectrum}.
\end{center}

For evaluating the matrix element we use the approximation
\begin{displaymath}
V_c,V_v\ll E_g,
\end{displaymath}
where $V_c$ and $V_v$ are the barrier heights for electrons and holes
respectively.
Obviously, this approximation also implies that $k_4^2+q^2\gg k_1^2$,
i.e. the total momentum of the excited electron is much larger than
that of the localized one. Integral over $x_1$ coordinate can be found
by integrating by parts. The n-th antiderivative of the function
$\psi_4 e^{-qx}$ is:
\begin{displaymath}
F_4^n(q,x) =(-1)^n\frac{(e^{qx}\psi_4(x))^{(n)}}{(k_4^2+q^2)^n}e^{-2qx}.
\end{displaymath}
Then the approximate expression for the matrix element $M_I$ may be
obtained:
\begin{equation}
M\approx M^{(1)}+M^{(2)},
\end{equation}
where
\begin{eqnarray}
M^{(1)}&=&-\frac{4\pi e^2}{\kappa_0 (q^2+k_4^2)}
\left({\cal F}(a/2)\int_{-a/2}^{a/2}
e^{qx_2}\psi_3^\ast(x_2)\psi_2(x_2)dx_2-\right.\nonumber\\
&-&\left.{\cal F}(-a/2)\int_{-a/2}^{a/2}
e^{-qx_2}\psi_3^\ast(x_2)\psi_2(x_2)dx_2\right).
\end{eqnarray}
Here
\begin{displaymath}
{\cal F}(a/2)=e^{-qa/2}\psi_{4s}^\ast(a/2)\psi_{1s}(a/2)
\left({3V_c+V_v\over 4E_g}-{\kappa_0-\tilde\kappa_0\over \kappa_0+
\tilde\kappa_0}\right).
\end{displaymath}
Index $s$ in the $\psi_{4s}$ and $\psi_1{1s}$ implies
that only $s$--components of the wave functions are taken,
$\tilde\kappa_0$ is the dielectric constant in the barrier region.
\begin{eqnarray}
M^{(2)}=\frac{4\pi e^2}{\kappa_0 (q^2+k_4^2)}
\int_{-a/2}^{a/2} \psi_4^\ast(x)\psi_3^\ast(x)\psi_2(x)\psi_1(x)dx
\end{eqnarray}
Note, as the wave functions are spinors the components of $\psi_4^\ast$
should be multiplied by components of $\psi_1$ and vice versa the
components of $\psi_3^\ast$ should be multiplied by those of $\psi_2$.

In this way it appears that the matrix element of Auger transition
splits two parts. The first of them is related to the presence of
heterobundaries and the second one corresponds to the short range
Coulomb scattering. The latter can be easily understood as during
Auger transition a large momentum is transmitted from the localized
electron to the excited one and this is possible only if the
scattering particles find themselves very close to each other.
Note that both $M^{(1)}$ and $M^{(2)}$ and, consequently, the matrix element
$M$ itself are in fact thresholdless matrix elements. Indeed, they are
not subject to any restrictions imposed on the initial momenta of carriers,
$k_1$, $k_c$, $k_h$. However, the mechanisms responsible for the momentum
nonconservation $k_1+k_2\not = k_3+k_4$ in the components $M^{(1)}$
and $M^{(2)}$ are different. In $M^{(1)}$ the nonconservation is related to
carrier scattering at the heteroboundary, and the same mechanism gives rise
to a thresholdless Auger process in scattering on a single heterobarrier
~\cite{5}. The reason why the conservation law breaks down for $M^{(2)}$
is that the volume of integration with respect to $x$ is restricted
to the quantum well region, which results in the appearance of a
function of the type ${\textstyle\sin{ka/2}\over\textstyle k}$
instead of  $\delta (k)$. The physical meaning of the above distinctions
between the matrix elements $M^{(1)}$ and $M^{(2)}$ is that the
$M^{(1)}$ corresponds to the true thresholdless process whose origin
is related to momentum scattering by heterobarriers. The matrix
element $M^{(2)}$ corresponds to a quasi-threshold
process, and, at the quantum well width $a$ approaching infinity,
it transforms into the conventional threshold matrix element.

\begin{eqnarray}
M^{(1)}&\approx&\frac{8\pi e^2}{\kappa_0 (q^2+k_4^2)(q^2+k_3^2)}
\left({3V_c+V_v\over 4E_g}-{\kappa_0-\tilde\kappa_0\over
\kappa_0+\tilde\kappa_0}\right)\times\nonumber\\
&\times&\left(\psi_4^\ast(a/2)\psi_1(a/2)\right)
\left(\psi_3^\ast(a/2)\psi_2(a/2)\right)^\prime(1\pm e^{-qa}).
\end{eqnarray}
Sign $\pm$ in the last brackets is chosen according to parity of the
production $\psi_3^\ast(x)\psi_2(x)$, $+$ corresponds to the even
production and $-$ corresponds to the odd one. In the case $qa\gg 1$
this exponent may be omitted and the matrix element $M^{(1)}$ corresponds
to the independent scattering at two heterobundaries.
Note, that the matrix element
$M$ equals to zero if the parities of productions  $\psi_3^\ast(x)\psi_2(x)$
and $\psi_4^\ast(x)\psi_1(x)$ are different.

Let us now analyze $M^{(2)}$. The integral entering into
$M^{(2)}$ is proportional
to the sum
\begin{equation}
\int_0^a
\psi_4^\ast(x)\psi_3^\ast(x)\psi_1(x)\psi_2(x)\,dx\propto
\sum\pm
{\sin{(k_4-k)}a/2\over k_4-k},
\label{eq:67}
\end{equation}
where $k$ runs through eight different values $k=\pm k_1\pm k_2\pm k_3$.
Of all terms in the sum (\ref{eq:67}) the largest is that for which
$k=k_1+k_2+k_3$.\footnote[2]{} The contributions to the sum
from other terms are less significant and will be neglected for the
sake of simplicity. Then the expression for the matrix elements of
the quasi-threshold Auger process takes the following form:

\begin{eqnarray}
M^{(2)}&&\approx
{\pi e^2\over \kappa_0 (q^2+k_4^2)}e^{i\delta}
{\hbar\gamma\over E_g}{1+2/3\alpha\over 1+\alpha}
A_cA_fA_cA_h{\sin{(k_f-k_{c1}-k_{c2}-k_h)a/2}\over k_f-k_{c1}-k_{c2}-k_h}
\times\nonumber\\
&&\times
\left\{
\begin{array}{c}
q_hk_ce^{i\phi_{2,3}}+q_ck_h,\\
q_cq_h\sin{\phi_{2,3}},
\end{array}
\right.
\;\left.
\begin{array}{c}
\nu_c =\pm\nu_h,\\
\nu_c=\mp\nu_h.
\end{array}
\right.
\label{eq:72}
\end{eqnarray}
Here $\delta$ is an insignificant phase coefficient, $A_i$ denotes
the normalizing constant, $\nu_c$ and $\nu_h$ are the spin indexes
introduced according to (\ref{eq:30}), $\phi_{2,3}$ is the angle between
lateral momenta of the electron and the hole. Note, that the matrix
element for transition into continuous spectrum ($M$) was split to
$M^{(1)}$ and $M^{(2)}$ in different way than in~\cite{JETP},
in the purpose to make the corresponding expressions more clear.

\begin{center}
{\it Calculation of the matrix element of Auger recombination for a
transition of an excited particle into the discrete spectrum}.
\end{center}

We now turn our attention to analyzing the matrix element of an Auger
transition in which the high-energy particle remains in bound state
$\psi_4$. This
case corresponds to the approximation $q_4\gg k_4$. The matrix element
can be calculated similarly as $M^{(2)}$ earlier:
\begin{equation}
M^{(3)}\approx
{4\pi e^2\over \kappa_0 (q^2+k_4^2)}
\int_{-a/2}^{a/2}
(\psi_4^\ast\psi_1)(\psi_3^\ast\psi_2)\, dx.
\label{eq:75}
\end{equation}

This integral can be readily calculated; however, the general formula is
rather cumbersome and will not be presented here. We shall only make an
estimate of $M_3$, valid in the case when bound carriers are in the ground
quantum state. Then we have:
\begin{equation}
M^{(3)}\approx
{1\over q^2+k_4^2}
e^{i\delta}
{\hbar\gamma\over Z}A_cA_fA_cA_ha/2\alpha q_cq_h\sin\phi_{2,3}\;\;
(\nu_c=-\nu_h).
\label{eq:77}
\end{equation}
Here $\alpha$ is a coefficient of order unity, resulting from integrating
the product of the envelopes of the carrier wave functions over the quantum
well:
\begin{equation}
\int_0^{a/2}
f_1f_2f_3f_4\, dx\approx
a/2\alpha,\;
\label{eq:76}
\end{equation}
where $f_i=\cos{k_ix}$, $i$-numerates the initial and final states of
particles taking part in the
AR process. For wide quantum wells, in which particles may occupy
different bound quantum states, $\alpha$ is given by (cf. (\ref{eq:67}):
\begin{equation}
\alpha={1\over 16}\sum_{\nu_1,\nu_2,\nu_3,\nu_4= 0,1}(-1)^{\nu_i\sigma_i}
\frac{sin{\left((-1)^{\nu_i}k_i\right)a/2}}{(-1)^{\nu_i}k_ia/2}
\end{equation}
Here by the index $i$ is meant summation from $1$ to $4$, and $\sigma_i$
characterizes the parity of the function $f_i$ ($\sigma_i=1$ and $0$ for
odd and even functions, respectively).

\begin{center}
\section*{\normalsize{\bf IV Auger recombination coefficient}}
\end{center}

To calculate the rate of AR, the probabilities of Auger
transition per unit time (\ref{eq:50}) should be summed over all
initial and final states of carriers with appropriate
weights--occupation numbers.
\begin{equation}
G=
{2\pi\over\hbar}
\sum_{{\bf k}_1,{\bf k}_2,{\bf k}_3,{\bf k}_4}
\langle
M^2
\rangle\cdot
f_1f_2(1-f_3)(1-f_4)
\delta (E_3+E_4-E_1-E_2).
\label{eq:78}
\end{equation}
Here $f_1$ and $f_2$ are the occupancies of the initial
states and $f_3$ and $f_4$ are those of final states,
\begin{displaymath}
\langle M^2\rangle=\sum_{\nu_1,\nu_2,\nu_3,\nu_4}|M_{fi}|^2
\end{displaymath}
is the squared Auger matrix element,
summarized over spins of initial and final states. It is
more convenient to choose electrons and holes as carriers
for the CHCC and CHHS processes, respectively. For high-excited
states the distribution function $f_4$ may be set
zero. Note that instead of $1-f_3$, we may write $\widetilde{f_3}$,
where $\widetilde{f_3}$ is the distribution function for carriers of the
opposite sign: holes for the CHCC process and electrons for
the CHHS process.

Contributions to the rate of Auger recombination from the
matrix elements  $M^{(1)},\,M^{(2)}$ and $M^{(3)}$ can be
separated since the excited particles occupy different quantum states.
The matrix elements $M^{(1)}$ and $M^{(2)}$, on the one hand, and
$M^{(3)}$, on the other, describe transitions in which the excited
particle occupies a state of continuous and discrete spectrum
respectively. It is more difficult
to separate the contributions from $M^{(1)}$ and $M^{(2)}$. Even though the
physical difference between these terms is preserved, there exists
a term of interference between them. At small quantum well widths
the interference is essential, since both processes are in fact
thresholdless; however, even when the interference is neglected
we still obtain a result of correct order, reflecting all the main
specific features of the AR coefficient as a function of temperature
and parameters of a structure with a quantum well. For a sufficiently
wide quantum well the interference between $M^{(1)}$ and $M^{(2)}$ may be
neglected. Indeed, while $M^{(1)}$ as a function of quasi-momentum shows no
extremum, the modulus of $M^{(2)}$ exhibits a maximum at the point
$k_4(q)+k_3=k_1+k_2$. When the quantum well width tends to infinity,
the maximum at this point is of the form of a $\delta$-function. In
accordance with the aforesaid, the AR probability for wide quantum wells,
corresponding to the matrix element $M^{(2)}$, has
a maximum (as a function of the longitudinal momentum of the
heavy hole $q_h$) at higher $q_h$ values than the probability
associated with $M^{(1)}$. With decreasing quantum well width
the maxima of these probabilities approach each other,
and the region of overlapping between these matrix
elements becomes larger.

The probabilities of an AR transition within the CHCC
process, corresponding to the matrix elements $M^{(1)}$ and $M^{(2)}$,
are shown in Fig. 1 as functions of the longitudinal
momentum of the heavy hole at different quantum well widths.
It can be seen that the interference between the
thresholdless process represented by $M^{(1)}$ and the quasi-threshold
process represented by $M^{(2)}$ is observed, in
accordance with the aforesaid, only for narrow quantum
wells. It should be noted that the AR probabilities are
rather smooth functions of the longitudinal momentum of
heavy hole ($q$), since, in calculating them, summation is taken
over discrete quantum states of carriers. At $q$
close to the maximum value determined by the conservation
of longitudinal momentum and energy, the AR
probability shows a square-root divergence eliminated
upon integration with respect to $q$, i.e., in calculating
the rate of AR. The probability of Auger transition for
the CHHS process has a form similar to that for the CHCC
process.

In line with the aforesaid, let us present the rate of AR
as follows
\begin{equation}
G=G_1+G_2+G_3,
\label{eq:79a}
\end{equation}
where the rate $G_1$ corresponds to a thresholdless Auger
process with the matrix element $M_1$, rate $G_2$ to a quasi-threshold
Auger process with the matrix element $M_2$, and rate $G_3$ to a
threshold Auger process with the matrix element $M_3$.

The expressions for the rates $G_1$ and $G_2$ can be derived
from (\ref{eq:78}) by changing in it summation over $k_4$ by
integration and passing from $\delta$-function with respect
to energy to $\delta$-function with respect to momentum. In what
follows we shall study the AR coefficients $C$ related to
the rate $G$ by
\begin{displaymath}
G=Cn^2p\;\;\mbox{and}\;\;G=Cp^2n
\end{displaymath}
for the CHCC and CHHS Auger processes respectively.
Here $n$ and $p$ are the 2D
densities of electrons and holes, respectively. For the CHCC
process we have:
\begin{eqnarray}
C_1&\approx&\frac{32\pi^2 e^4}{\kappa_0^2}\frac{\hbar\gamma^2}{E_g^3}
{F(\Delta_{so}/E_g)\over
a(a+2/\kappa_c)^2}\frac{k_c^2\kappa_c^2}{(k_c^2+\kappa_c^2)
^2}{V_c\over E_g}\left( {3V_c+V_v\over 4E_g}
-{\kappa_0-\tilde\kappa_0\over \kappa_0+\tilde\kappa_0}\right)^2
\left< \frac{q_h^2k_h^2}{(q_h^2+k_h^2)^3}{1\over k_4(q_h)}\right>,
\label{eq:80a}
\end{eqnarray}
where
\begin{displaymath}
F(x)=\left({1+2x/3\over 1+x}\right)^2
{1+7x/9+x^2/6\over (1+x/2)(1+4x/9)}
\;\mbox{is a coefficient of order unity},
\end{displaymath}
Note, that if Kane's parameter $\gamma$ is discontinuous
than the term
${E_{0c}\over 2E_g}\left({\delta\gamma\over \gamma}\right)^2$
should be added to
$\left({3V_c+V_v\over 4E_g}-{\kappa_0-\tilde\kappa_0\over
\kappa_0+\tilde\kappa_0}\right)^2$, where $E_{0c}$ is the
electron size quantization energy and $\delta\gamma=\gamma
-\tilde\gamma$ is the difference between Kane's parameters in
the quantum well and barrier region. However, this addition is
usually negligible.
The angular brackets in (\ref{eq:80a}) and later
denote averaging over the heavy hole distribution
function. In the case of Boltzmann distribution, which
is commonly the case for holes, this averaging looks like:
\begin{eqnarray*}
\langle f(q_h,k_h)\rangle&=&{1\over Z}\sum_n\int_0^\infty q_hf(q_h,k_{hn})
e^{-{k_{hn}^2+q_h^2\over q_T^2}},\;\; \mbox{where}\\
Z&=&{2\over q_T^2}\sum_n e^{-{k_{hn}^2\over q_T^2}},
\end{eqnarray*}
$q_T=\sqrt{2m_hT}/\hbar$ is the thermal wave vector of heavy holes,
$k_{hn}$ is the wave vector corresponding to the n-th size quantization
level of heavy holes.

For the CHHS process the following expression for $C_1$ can be
derived:
\begin{eqnarray}
C_1\approx\frac{2\pi^2 e^4}{\kappa_0^2\hbar}{V_c\over E_g}\frac{k_c^2
\kappa_c^2}{(k_c^2+\kappa_c^2)^2}{\tilde F(\Delta_{so}/E_g)
\over a^2(a+2/\kappa_c)}{\hbar^3\over m_{so}^3(E_g-\Delta_{so})^3}
\left<\frac{k_{h1}^2k_{h2}^2q_{h1}^2(q_{h1}^2+q_{h2}^2)}
{(q_{h1}^2+k_{h1}^2)^3(q_{h2}^2+k_{h2}^2)}\right>, where
\label{x}
\end{eqnarray}
\begin{displaymath}
\tilde F(x)=\frac{\left(2x+3(1-x)(1-m_{so}/m_h)\right)^2}
{2x^2+\left(x+3(1-x)(1-m_{so}/m_h)\right)^2}\frac{1+2x/3}{1+x}.
\end{displaymath}
In the letter case averaging over distribution functions of two
holes occurs. In deriving (\ref{x}) we assumed that
$E_g-\Delta_{so}\gg m_h/m_{so} T$.

Similarly, we obtain $C_2$ for the CHCC process:
\begin{eqnarray}
C_2\approx\frac{\pi^2 e^4}{\kappa_0^2}{\hbar^3\gamma^4\over E_g^5}
{F(\Delta_{so}/E_g)\over a(a+2/\kappa_c)^2}
\left<\frac{q_c^2k_h^2+q_h^2(k_c^2+{1\over 2}q_c^2)}
{(q_h^2+k_h^2)k_4(q_h)}
\frac{1-cos(k_f-k_h-2k_c)a}{2(k_f-k_h-2k_c)^2}\right>.
\label{eq:81a}
\end{eqnarray}
Direct calculation of the Auger coefficient $C_2$ for the
CHHS process gives a cumbersome result. We  present here
a simplified expression valid for sufficiently narrow
quantum wells at $k_c\gg q_c$.
\begin{eqnarray}
C_2&\approx&\frac{\pi^2e^4}{4\kappa_0^2}{E_c\over E_g}{\hbar^3\over
m_{so}^2(E_g-\Delta)^3}{\tilde F(\Delta_{so}/E_g)\over a^2(a+2/\kappa_c)}
\left<\frac{1-cos(k_{so}-k_{h1}-k_{h2}-k_c)a}
{2(k_{so}-k_{h1}-k_{h2}-k_c)^2}\right.
\times\nonumber\\
&\times&\left.\frac{q_{h2}^2\left((k_{so}^2+k_{h1}^2)q_{h1}^2
+q_{h2}^2k_{h1}^2+2k_{h1}^2({\bf q}_{h1}{\bf q}_{h2})
+[{\bf q}_{h1}\times{\bf q}_{h2}]^2\right)}
{(q_{h1}^2+k_{h1}^2)(q_{h2}^2+k_{h2}^2)k_{so}}
\right>
\end{eqnarray}
And, finally, we have for $C_3$ for the CHCC process:
\begin{eqnarray}
C_3\approx\frac{32\pi^2 e^4}{\kappa_0^2\hbar E_g}{a\over(a+1/\kappa_c)^3}
\frac{1+{7\over 9}x+{1\over 6}x^2}{(1+x/3)^2}\,\frac{1+{2\over 3}x}{1+x}
\cdot\left<\frac{q_{th}^2}{q_T^2}\frac{q_c^2}{(q_{th}^2+k_h^2)^3}
e^{-\frac{q_{th}^2}{q_T^2}}\alpha^2\right>_n.
\label{eq:82a}
\end{eqnarray}
Here $x=\Delta_{so}/E_g$, $\alpha$ is a multiplier introduced in
(\ref{eq:77}).
In the last case we average only over discrete quantum
states of heavy holes. The threshold
momentum $q_{th}$ is found from the conservation law for the
energy and the longitudinal component of momentum:
\begin{displaymath}
E_f(\sqrt{k_f^2+q_{th}^2})=E_g+\frac{\hbar^2(q_{th}^2
+k_h^2)}{2m_h}+\frac{\hbar^2(k_{c1}^2+k_{c2}^2)}{2m_c}.
\end{displaymath}
For simlicity we neglected here by longitudinal momenta  of electrons,
because they are  small. However, we took into account
the size quantization energies of electrons as they change the
effective band gap in the quantum well.
Expanding the energy of excited electron $E_f$ into a series
in terms of momenta in the vicinity of $q_{th}=Q$, where $Q$ is the
electron momentum corresponding to an energy equal to $E_g$
($\textstyle Q\approx\sqrt{\frac{4m_cE_g}{\hbar^2}}$), we get the
following estimation for the threshold momentum:
\begin{equation}
q_{th}\approx\sqrt{\frac{4m_cE_g}{\hbar^2}+{3\over 2}\left(k_c^2
+{m_c\over m_h}k_h^2\right)}.
\label{eq:83}
\end{equation}
If the quantum well width tends to infinity, the
threshold momentum approaches its bulk value~\cite{2}.
Account must be taken of the fact that for wide quantum
wells with large number of levels the introduced
multiplier $\alpha$ (see eq. (\ref{eq:77})) tends to a $\delta$-function
expressing the conservation law for the transverse quasi-momentum
component:
\begin{displaymath}
\alpha^2\longrightarrow{\pi\over 128}a\sum\delta(k_h\pm k_{c1}\pm k_{c2}
\pm k_{c4}).
\end{displaymath}
For large quantum well widths, provided that the condition
$V_c\ll E_g$ holds, the inequality $C_3\ll C_2$ is valid,
since the ratio $C_3/C_2\approx\sqrt{V_c/E_g}$. Hence, for
wide wells $C_3$ may be neglected as compared with $C_2$.
If $V_c\lesssim E_g$, the following relation holds
for wide quantum wells
$C_3/C_2\approx \sqrt{V_c/(E_g-V_c)}\geq 1$. For narrow
quantum wells the threshold energy of the CHCC process
increases (see eq. (\ref{eq:83})) and the AR coefficient (\ref{eq:82a})
decreases relative to the bulk value by a factor
\begin{displaymath}
e^{3k_c^2\over 2q_T^2}\approx e^{{3m_c\over
2m_h}{E_{0c}\over T}}.
\end{displaymath}
The characteristic width of a quantum well for which this
phenomenon becomes essential can be readily evaluated
from the condition that the exponent is unity:
\begin{equation} E_{0c}\approx
T {2m_h\over 3m_c}\Leftrightarrow a\approx \pi{1\over q_T}.
\label{eq:83'}
\end{equation}
Thus, at quantum well widths $a$ less than several
reciprocal thermal momenta $a\lesssim\pi q_T$ the
threshold energy $E_{th}^3(a)$ becomes much higher than the
bulk value $E_{th}^{bulk}$ (see Fig. 2). For semiconductor
compounds A$_{III}$B$_V$ the equality
(\ref{eq:83'}) is fulfilled at room temperature at a quantum
well width of order of $100$\AA.

For the threshold CHHS process the heavy hole momenta
are not specified by the threshold conditions and, therefore,
we have to perform integration with respect to them. It
seems impossible to derive analytically the exact result
for the Auger coefficient $C_3$ for the CHHS process in view
of the fact that the matrix element $M_3$ is rather
cumbersome. However, an approximate
expression can readily be obtained by factoring out the
averaged squared matrix element from the integrand sign:
\begin{eqnarray}
C_3&&\approx{2\pi\over\hbar}\langle M_3^2\rangle{1\over 2\pi^2q_T^4}
\int q_{h1}dq_{h1}q_{h2}dq_{h2}d\phi_{h1}d\phi_{h2}e^{-{q_{h1}^2+q_{h2}^2
\over q_T^2}}\times\nonumber\\
&&\times\delta\left(\tilde{E_g}-\Delta-
\frac{\hbar^2({\bf q_{h1}+q_{h2}})^2}{2m_{so}}+{\hbar^2q_{h1}^2\over
2m_h}+{\hbar^2q_{h2}^2\over 2m_h}\right),
\end{eqnarray}
Here $\tilde E_g=E_g+E_{0c}+2E_{0h}-E_{0\,so}$, where
$E_{0c},\;E_{0h}\;E_{0\,so}$ are the energies of
size quantization for electrons, holes, and SO holes,
respectively. Let us introduce a threshold momentum,
setting it equal to
\begin{displaymath}
Q_{th}^2=\frac{2(\tilde{E_g}-\Delta)m_{so}}{\hbar^2(2-\mu_{so})},
\end{displaymath}
where $\mu_{so}=m_{so}/m_h$.
Then the expression for $C_3$ takes the form
\begin{equation}
C_3\approx{2m_{so}\over \hbar^3Q_{th}^2}e^{-{Q_{th}^2\over q_T^2}}
\langle M_3^2\rangle
\end{equation}
Taking into account that $Q_{th}\gg k_{so}$, we obtain:
\begin{equation}
C_3\approx{\textstyle 256m_{so}\pi^2e^4\over \textstyle \hbar^3\kappa_0^2}
\frac{Q_{th}^2}{(Q_{th}^2+2k_h^2)^4}{\textstyle V_c\over \textstyle E_g}
\frac{k_c^2}{k_c^2+\kappa_c^2}\frac{(1-\lambda_{so})^2}{1+2\lambda_{so}^2}
\tilde\alpha^2e^{-{Q_{th}^2\over q_T^2}},
\end{equation}
where $\tilde\alpha$ is a multiplier defined similarly as for CHCC
process (see eq. (\ref{eq:77})), $\lambda_{so}$
is derived from $\lambda_l$ (see eq. (\ref{eq:34})) by substitution
$k_{so}$ instead of $k_l$.
Let us consider in more detail the Auger recombination
coefficient $C_2$ for the quasi-threshold CHCC process.
At $a\longrightarrow\infty$ a substitution can be made in the
function being averaged in (\ref{eq:81a}):
\begin{equation}
\frac{1-\cos{(k_f-k_h-2k_c)a}}{2(k_f-k_h-2k_c)^2}\longrightarrow
\frac{\pi a}{2}\delta(k_f-k_h-2k_c).
\label{eq:84a}
\end{equation}
This formula clearly shows the occurrence of a threshold
in this limit, and the coefficient $C_2$ transforms into a
3D expression on being multiplied by $a^2$.
For comparison we present both the result of Gelmont~\cite{2} for
$C_{3D}$ and our limiting expression.
\begin{equation}
C_{3D}=6\sqrt{2\pi^5}\frac{e^4 m_c\hbar^3}{\kappa_0^2}\frac{1}{E_g^{5/2}
T^{1/2}m_c^{1/2}m_h^{3/2}}e^{-{2m_c\over m_h}{E_g\over T}},
\label{eq:84}
\end{equation}
\begin{equation}
C_2\cdot a^2=6{16\sqrt{2\pi^5}\over 27}\frac{e^4 m_c\hbar^3}{\kappa_0^2}
\frac{1}{E_g^{5/2} T^{1/2}m_c^{1/2}m_h^{3/2}}
e^{-{2m_c\over m_h}{E_g\over T}}.
\label{eq:85}
\end{equation}
The factor 4 in (\ref{eq:85}) results from the necessity to take
into account, in calculating $M_2$ in accordance with (\ref{eq:72}),
not only the terms with $k=k_{c1}+k_{c_2}+k_h$, but also those with
$k=k_{c1}-k_{c_2}+k_h$, $k=-k_{c1}+k_{c_2}+k_h$, and
$k=-k_{c1}-k_{c_2}+k_h$. When the quantum well width tends to
infinity, all the four terms make equal contributions to
$C_2$. As can be seen, the only difference
between the expressions (\ref{eq:84}) and (\ref{eq:85})
is in a numerical coefficient. A small discrepancy of $\approx2/3$ times
is due to the necessity to distinguish between the size quantization
momenta of the two localized electrons: $k_{c1}\not =k_{c2}$, besides
the expression (\ref{eq:84}) was derived when $\Delta_{so}$ tends
to infinity, while obtaining the expression (\ref{eq:85}) it was
assumed that $\Delta_{so}\lesssim E_g$. In calculating
(\ref{eq:85}) we neglected the quantity $V_c$ as compared with $E_g$.
In the general case, $(C_2+C_3)\,a^2$ should be written instead of
$C_2\,a^2$ to make expression (\ref{eq:85}) valid. However, the limiting
transition from the quasi-threshold to the threshold Auger process
(see eq. (\ref{eq:84a})) can be realized only for very large quantum wells.
Analyzing the probability of Auger transition as a function of the heavy
hole momentum, one can obtain a qualitative criterion for this transition.
As mentioned above, the quasi-threshold Auger process dominates in
wide quantum wells. The probability of this process has two characteristic
extrema (see eq. (\ref{eq:81a})). The first of them corresponds to the maximum of
squared Auger transition matrix element in the vicinity of the threshold
momentum. The width of this maximum is of the order of the inverse quantum
well width. The second extremum lies in the vicinity of the thermal momentum
of heavy holes. The Auger coefficient $C_2$ can be estimated as a
sum of the probabilities taken in these extrema, multiplied by
the corresponding widths. Then
\begin{equation}
C_2\approx C_2^{th}(Q_h\approx q_{th})+C_2^{T}(Q_h\approx q_T),
\end{equation}
where $Q_h$ is the value of the heavy hole momentum: $Q_h^2=k_h^2+q_h^2$;
\begin{eqnarray}
{C_2^T\over C_2^{th}}\approx {\lambda_{E_g}\over a}
\left({T\over E_{th}}\right)^{3/2}e^{E_{th}\over T}.
\end{eqnarray}
Here $\lambda_{E_g}\approx 2\pi/q_{th}$ is a characteristic wave
length of an electron having energy close to $E_g$.
Comparing the terms $C_2^{th}$ and $C_2^T$ one can obtain the criterion
of the transition from the quasi-threshold to the threshold Auger process:
\begin{equation}
a\gg a_c,\;\;{where} a_c=\lambda_{E_g}\left({T\over E_{th}}\right)^{3/2}
e^{E_{th}\over T}.
\end{equation}
For semiconductors with an energy gap of the order of $1\,eV$ the
critical width ($a_c$) may be as large as several thousand Angstroms.
However, the value $a_c$ is considerably larger than the free path length of
carriers in semiconductors. This obviously shows that the correct
derivation of Auger rate in homogeneous semiconductors should involve
momentum scattering process if the critical width $a_c$ there
exceeds the free path length~\cite{22}.

With decreasing quantum well width, the maximum of the
probability $w_2$ as a function of the heavy hole momentum
shifts to lower values (see Fig. 1). This reduces
the threshold energy of the process, and, as a
consequence, makes weaker the temperature dependence of
the AR coefficient.

Figure 2 shows the threshold energy
of the CHCC process as a function of quantum well width
for all the three mechanisms of Auger recombination $C_1$,
$C_2$, and $C_3$ separately and for the overall process
$C=C_1+C_2+C_3$, found from the formula:
\begin{equation}
E_{th}^i(T)=T^2 \frac{d\ln{C_i}}{dT},\;\;i=1,2,3
\label{eq:85'}
\end{equation}
The threshold energy for the quasi-threshold Auger process
(see Fig. 2) is less than its bulk value. The reason is that the
critical quantum well width $a_c\approx 2000\,\overcirc{A}$ is greater
than the maximum width shown in the figure. The threshold energy for
the thresholdless Auger process decreases with quantum well width and
becomes negative at a certain thickness. This is due to the fact that
the Auger coefficient $C_1$ decreases with increasing
temperature for wide enough
quantum wells (see Fig. 5). With increasing quantum
well width, the total threshold energy tends to its limiting value
$E_{th}^{bulk}$ denoted in the figure.

We now turn our attention to the thresholdless Auger
process. As already noted, the probability of a
thresholdless Auger transition shows no extrema as a
function of the heavy hole momentum. Therefore, the
coefficient $C_1$ has a weak nonexponential temperature
dependence. This phenomenon was first studied by Zegrya
and Kharchenko~\cite{5}. In addition, the function $C_1(T)$ is
nonmonotonic and has a maximum. The presence of this
maximum can readily be explained. At low temperature and,
correspondingly, small longitudinal momenta of carriers
their wave functions are nearly orthogonal and the $C_1$
value is small. With increasing temperature, the
characteristic momentum transferred in Coulomb
interaction (approximately equal to the thermal momentum
of heavy hole) grows. This is the reason why at low
temperature the Auger coefficient increases with
temperature. As the temperature is elevated further, the
AR coefficient $C_1$ passes through a maximum and starts to
decrease, since the long-range Coulomb interaction
responsible for the Auger process is low for large
transferred momenta. The temperature at which the maximum
occurs can readily be evaluated by equating the energy of
size quantization of holes to temperature:
$\textstyle T\approx \frac{\hbar^2\pi^2}{2m_ha^2}$. Note that
there would be no such maximum if the overlap integral
$I_{ch}$ were taken to be proportional to the momentum
transferred. Such an assumption, having in our opinion no
justification for the majority of structures
investigated, is frequently used
(see, e.g., ref.{\normalsize \ref{eq:20}})
and gives incorrect expressions for the rate of AR and
incorrect dependence of this quantity on temperature and
quantum well parameters.

The AR coefficient $C_1$ depends rather strongly on the
quantum well width $a$. Depending on which term
predominates in (\ref{eq:80a}), the coefficient $C_1$ decreases with
increasing $a$ either as $1/a^3$ or as $1/a^5$ or as $1/a^7$. In any
case, even being multiplied by $a^2$, $C_1$ remains a
decreasing function of quantum well width. For this
reason such a process may be dominant only for
sufficiently narrow quantum wells. At $\textstyle a\approx 1/\kappa_c$ the
coefficient $C_1$ exhibits a maximum related to the weak
overlapping of carrier wave functions. With the quantum
well width decreasing further, the rate of the
thresholdless Auger process falls gradually. The
similar expression for $C_1$ in the CHCC process was obtained
by Dyakonov and Kachorovskii, and Zegrya et al.~\cite{11,12}.

Figure 3 shows the AR coefficients $C_1$ and $C_2$ for the CHCC
process as functions of quantum well width at different
temperatures for a model structure based on InGaAsP. It
can be seen that, firstly, both $C_1$ and $C_2$ show a sharply
pronounced maximum. Interestingly, the positions of these
maxima are practically temperature-independent. Secondly,
the relative role of the quasi-threshold process becomes more
important with increasing temperature. The threshold
process is not shown in the figure, since at the quantum
well widths considered the coefficient $C_3$ is much smaller
(by several orders of magnitude) than $C_1$ and $C_2$. For this
reason the dependence of $C_3$ on quantum well width is
shown separately in Fig. 4. Note that for this process
the maximum is achieved at a much wider quantum well than
for the quasi-threshold or thresholdless process. This is
in the first place due to the reduction of the threshold
energy of the threshold process with increasing quantum
well width (Fig. 2), rather than to the overlapping of
the wave functions. The coefficients $C_1$ and $C_2$ as
functions of quantum well width for the CHHS process
qualitatively agree with the curves presented in Fig. 3.

Figure 5 shows the temperature dependence of the overall
AR coefficient for the CHCC process and the partial
contributions from the thresholdless and quasi-threshold
mechanisms at different quantum well widths. It can be
seen that at low temperature and sufficiently wide
quantum wells the thresholdless Auger process
predominates ($C_1>C_2$), and at high temperature, conversely,
the quasi-threshold process becomes more important ($C_2>C_1$).
Therefore, the curve describing the temperature
dependence of the overall AR coefficient has a
characteristic shape with a maximum and a minimum. With
increasing quantum well width, both the maximum and the
minimum of the AR coefficient shift to lower temperature
and, in the limit of infinitely wide quantum well,
disappear. Thus, in the case of a homogeneous
semiconductor the AR coefficient is a monotonic function
of temperature. Note that the Boltzmann distribution
of carriers was used in calculating the Auger coefficients
as functions of temperature. At low temperature
both electrons and holes are as a rule described by Fermi-Dirac
distribution function. Thus, the average momenta of particles
participating in the Auger transition depend on temperature only
slightly. As a result,
at low temperatures the Auger coefficient is a smoother function
of temperature than in the case of the Boltzmann statistic
and it does not tend to zero at $T\rightarrow 0$. Figure 6 shows
the thresholdless Auger coefficient $C_1$ versus
temperature at various Fermi energies for quantum wells with
different widths. Essential discrepancies between the values of
the Auger coefficients for the Fermi-Dirac and Boltzmann distributions
take place only in the case $T\ll E_F$, where
$E_F$ is the Fermi energy for holes. This condition can be
realized only at very low temperature where the Auger process ceases
to be an important mechanism of recombination.

\begin{center}
\section*{\normalsize\bf V Phonon-assisted Auger
recombination in quantum wells}
\end{center}
At high temperature the threshold AR process predominates
in the homogeneous semiconductor ($\textstyle C_{3D}
\propto e^{-{E_{th}\over T}}$). However, at sufficiently low
temperature such a process becomes exponentially weak.
In this case the rate of the Auger process is no longer determined
by the scattering of two free electrons. Mechanisms leading to threshold
removal are to be taken into account. It is commonly
believed that the primary mechanism of this kind is
emission or absorption of a virtual optical phonon. At
the expense of a large momentum transferred to the
phonon, the AR threshold for heavy holes is removed, and
the rate of such an Auger process is a power-law function
of temperature~\cite{3,4}$^,$\cite{23}. The probability of
the phonon-assisted AR is calculated in terms of the second-order
perturbation theory in electron-electron and electron-
phonon interaction~\cite{21}. However, the possibility of
removing the threshold through interaction with phonons
is not the only one. At high hole concentrations, hole-hole
scattering may become a more effective mechanism of
removing the threshold for the Auger process. This is why
the question of the AR mechanism in homogeneous
semiconductors at low temperature still remains open. We
shall discuss this problem in more detail elsewhere~\cite{22}.
In the present work we follow the commonly accepted viewpoint
that there exists a competition between the
phonon-assisted and direct AR processes. For quantum
wells the situation differs strongly from the 3D case,
owing to the presence of a thresholdless process even in
the first order of the perturbation theory. Therefore, it is
{\it a priori} evident that the conditions under which the
phonon-assisted AR dominates over the direct process
depend strongly on quantum well width.

As already noted, in the 2D case there exist three AR
processes: threshold, quasi-threshold, and thresholdless.
It will be recalled that in sufficiently narrow quantum
wells thresholdless and quasi-threshold AR processes are
predominant at all temperatures.
Below, it will be shown that the electron-phonon
interaction may affect significantly the rate of AR at
low temperature. The AR coefficient for a phonon-assisted
CHCC process with a threshold matrix element of
electron-electron interaction is comparatively easily calculated
when~\cite{23}
\begin{equation}
E_g\gg 2\mu E_g\gg \hbar\omega_{ LO},T
\label{eq:86}
\end{equation}
where $\textstyle \omega_{LO}$ is the optical phonon frequency,
$\textstyle\mu=m_c/m_h$.
It can be shown that the coefficient of phonon-assisted AR is
related to the previously calculated AR coefficient (\ref{eq:85}) by
\begin{eqnarray}
C^3_{ph}\approx C_3\frac{e^2\hbar\omega_{ LO}}{2\overline\kappa a}\frac{T}
{E_{th}^{2D}}g(a,k_{th})\frac{1}{e^{\hbar\omega_{ LO}\over T}-1}\left[
\frac{e^{\hbar\omega_{ LO}\over T}}{\left( E_{th}^{2D}-\hbar
\omega_{ LO}\right)^2}+
\frac{1}{\left( E_{th}^{2D}+\hbar\omega_{ LO}\right)^2}\right]
e^{E_{th}^{2D}\over T},
\label{eq:87}
\end{eqnarray}
where $\textstyle \overline\kappa={\kappa_0\kappa_\infty\over \kappa_0
-\kappa_\infty}$, $\kappa_\infty$ is the high-frequency permittivity of the
medium, $g(a,k_{th})$ is a factor reflecting the 2D character of
holes~\cite{24}
\begin{equation}
g(a,k_{th})=k_{th}^2a^2 \left[ {1\over 2k_{th}^2a^2}
+{1\over 4(\pi^2+k_{th}^2a^2)}\right]\,\left[1-\frac{1-e^{-2k_{th}a}}
{2k_{th}a}\frac{2\pi^4}{(\pi^2+k_{th}^2a^2)(2\pi^2+3k_{th}^2a^2)}\right].
\label{eq:88}
\end{equation}
For comparison, we present an AR coefficient calculated
for the phonon-assisted 3D case
\begin{eqnarray}
C^{3D}_{ph}\approx C^{3D} \frac{e^2\hbar\omega_{ LO}}{2\sqrt\pi
\overline\kappa}\left(\frac{T}{E_{th}^{3D}}\right)^{3/2}
\frac{k_{th}}{e^{\hbar\omega_{ LO}\over T}-1}\left[
\frac{e^{\hbar\omega_{ LO}\over T}}{\left( E_{th}^{3D}
-\hbar\omega_{LO}\right)^2}+\frac{1}{\left( E_{th}^{3D}
+\hbar\omega_{ LO}\right)^2}\right] e^{E_{th}^{3D}\over T}.
\label{eq:89}
\end{eqnarray}
It can be seen that the results for 2D and 3D phonon-assisted
AR processes with threshold matrix elements of
electron-electron interaction are closely allied. A
significant difference for the case of narrow quantum
wells is that the threshold energy $E_{th}^{2D}$ increases owing
to the presence of carrier size quantization levels (see eq. (\ref{eq:83})).
Correspondingly, the criterion for predominance of
the phonon-assisted AR process ($C_3^{ph}$) over the direct
threshold Auger process ($C_3$) in quantum wells is met at
somewhat higher temperature than in the 3D case. However,
as already noted (see Section III), the rate of
the threshold Auger process in narrow quantum wells is in itself by
several orders of magnitude lower than those of
the thresholdless and quasi-threshold processes ($C_3\ll(C_1,\,C_2)$).
Hence, the phonon-assisted AR process with a
threshold matrix element of electron-electron interaction
cannot compete with thresholdless and quasi-threshold
processes either ($C^3_{ph}\ll C_1,\,C_2$).

Let us now consider a phonon-assisted Auger process with
the thresholdless matrix element ($M_{ee}$) for the CHCC
process. Direct calculation of the probability of this
process results in a singularity at the energy of a hole in the
virtual state energy equal to the sum of the energies of the optical phonon
and the hole in the final state:
\begin{equation}
w_{i\rightarrow f}=\pm\frac{2\pi}{\hbar}\sum_s\frac{\left|M_{ee}\right|^2
\left|M_{hp}\right|^2}{(E_{s}\mp \hbar\omega_{LO} -E_h)^2}\frac{e^{\pm
\frac{\hbar\omega_{LO}}{T}}}{e^{\pm\frac{\hbar\omega_{LO}}{T}}-1}
\delta(E_i-E_f)d\nu_f,
\label{eq:89a}
\end{equation}
where $E_s$ is the energy of the virtual hole, $and M_{ep}$ is the matrix
element of scattering of the virtual hole by an optical
phonon, with the signs plus and minus corresponding to
phonon emission and absorption, respectively. Note that
in the 3D case the summation over intermediate states is
in fact reduced to taking a single summand. In the 2D
case, nothing of the kind occurs, and the sum is taken
over discrete quantum states. To eliminate  the
divergence in the denominator of the expression (\ref{eq:89a}),
account must be taken of transitions not into stationary,
but quasi-stationary states, i.e., states with complex
energy. In this case the pole (\ref{eq:89a}) will transform into a
region of complex energy values:
\begin{displaymath}
w_{i\rightarrow f}\propto{1\over(E_{s}\mp\hbar\omega_{LO}-E_h)^2+\Gamma^2},
\end{displaymath}
where $\textstyle\Gamma=\hbar/\tau$.
The characteristic lifetimes $\tau$ corresponding
to these states may vary over a wide range, depending on
temperature, free carrier density, etc. It only
makes sense to consider a resonant phonon-assisted
process in terms of the second-order perturbation theory when
the halfwidths of the quasi-stationary hole and the phonon states
are less than the energy of the optical phonon ($\hbar\omega_{LO}$).
Otherwise, the Auger coefficient must be calculated in
the first order of the perturbation theory, using the Lorentz
function ($\textstyle f(\Delta E)={1\over\pi}
{\Gamma\over\Delta E^2+\Gamma^2}$) instead of the $\delta$-function
expressing the energy conservation law. For a phonon-assisted AR process
with a quasi-threshold matrix element of electron-electron
interaction both the resonant and virtual Auger
processes are possible, with the former predominant in
narrow quantum wells and the latter in sufficiently wide
quantum wells.

In the general case the Auger coefficient for a phonon-assisted
process with the quasi-threshold and the thresholdless matrix elements
$M_{ee}$ may be written as
\begin{equation}
C_{ph}=C_{ph}^1+C_{ph}^2,
\label{eq:90}
\end{equation}
where
\begin{eqnarray}
C_{ph}^1=&&{\pi\omega e^2\over \overline\kappa Z}
\frac{e^{\hbar\omega_{ LO}/T}}{e^{\hbar\omega_{ LO}/T}-1}
\sum_{m,n,\nu_n}\int
\frac{d^2Q}{(2\pi)^2}\frac{d^2q_h}{(2\pi)^2}\left( \frac{\partial E_4}
{\partial k_4}\right)^{-1}\nonumber\\
&&\frac{\left|M_{ee}(n,{\bf q_h}+{\bf Q}\right|^2}
{\left({\hbar^2(n^2-m^2)\pi^2\over 2a^2m_h^2}+{\hbar^2({\bf q_h}
+{\bf Q})^2\over 2m_h}-{\hbar^2 q_h^2\over 2m_h}-\hbar\omega_{ LO}
\right)^2+\Gamma^2}J_{n,m}(Q)f_h(m,q_h)\\
\label{eq:91}
C_{ph}^2=&&{\pi\omega e^2\over \overline\kappa Z}
\frac{1}{e^{\hbar\omega_{ LO}/T}-1}
\sum_{m,n,\nu_n}\int
\frac{d^2Q}{(2\pi)^2}\frac{d^2q_h}{(2\pi)^2}\left( \frac{\partial E_4}
{\partial k_4}\right)^{-1}\nonumber\\
&&\frac{\left|M_{ee}(n,{\bf q_h}+{\bf Q}\right|^2}
{\left({\hbar^2(n^2-m^2)\pi^2\over 2a^2m_h^2}+{\hbar^2({\bf q_h}
+{\bf Q})^2\over 2m_h}-{\hbar^2 q_h^2\over 2m_h}+\hbar\omega_{ LO}
\right)^2+\Gamma^2}J_{n,m}(Q)f_h(m,q_h)\nonumber.
\end{eqnarray}
Here:
\begin{displaymath}
Z=\sum_m\int {d^2q_h\over (2\pi)^2}f_h(m,{\bf q_h}),
\end{displaymath}
\begin{eqnarray*}
&&J_{n,m}(Q)={a\over 2}\frac{(1+\delta_{m,n})
\left( (m+n)^2\pi^2+Q^2a^2\right)
+(m-n)^2\pi^2+Q^2a^2}{\left( (m+n)^2\pi^2+Q^2a^2\right)
\left( (m-n)^2\pi^2+Q^2a^2\right)}(1-\varepsilon),\\
&&\varepsilon=\frac{Qa\left(1-(-1)^{m+n}
e^{-Qa}\right)32\pi^4n^2m^2}{\left(
(m+n)^2\pi^2+Q^2a^2\right)\left((m-n)^2\pi^2+Q^2a^2\right)}\times\\
&&\times\frac{1}{(1+\delta_{m,n})\left((m+n)^2\pi^2+Q^2a^2\right)
+(m-n)^2\pi^2+Q^2a^2}.
\end{eqnarray*}

The function $J_{n,m}(Q)$ has been calculated for a nondegenerate band.
In the case of phonon scattering by heavy holes its value will be
somewhat lower. However, this fact is insignificant for our purposes.
For momenta of bound electrons in the matrix element of electron-electron
interaction in (\ref{eq:91}) should be substituted their mean thermal
values. The Auger coefficients $C_{ph}^1$ and $C_{ph}^2$
correspond, respectively, to phonon emission and absorption.
Irrespective of the type of the matrix element of Coulomb interaction,
the phonon-assisted Auger process is thresholdless. This corresponds
to the main contribution to the Auger coefficient $C_{ph}$
coming from the hole momenta of the same order of magnitude as the
thermal momentum. Therefore, in calculating $C_{ph}$ me may substitute
for the longitudinal hole momentum $q_h$ its mean thermal value.
This simplifies considerably the expression for the Auger coefficient:
\begin{eqnarray}
C_{ph}^1\approx &&{\pi\omega e^2\over
\overline\kappa} \frac{e^{\hbar\omega_{LO}/T}}
{e^{\hbar\omega_{ LO}/T}-1} \sum_{m,n,\nu_n}\int
\frac{d^2Q}{(2\pi)^2}\left< \left(\frac{\partial E_4}
{\partial k_4}\right)^{-1}\right>\nonumber\\
&&\frac{\left|M_{ee}(n,{\bf q_h}+{\bf Q}\right|^2}
{\left({\hbar^2(n^2-m^2)\pi^2\over 2a^2m_h^2}+{\hbar^2({\bf q_h}
+{\bf Q})^2\over 2m_h}-{\hbar^2 q_h^2\over2m_h}-\hbar\omega_{ LO}
\right)^2+\Gamma^2}J_{n,m}(Q)\frac{p_m}{\sum_ip_i},
\label{eq:92a}\\
C_{ph}^2\approx &&{\pi\omega e^2\over
\overline\kappa} \frac{1}{e^{\hbar\omega_{ LO}/T}-1}
\sum_{m,n,\nu_n}\int \frac{d^2Q}{(2\pi)^2}\left< \left(
\frac{\partial E_4}
{\partial k_4}\right)^{-1}\right>\nonumber\\
&&\frac{\left|M_{ee}(n,{\bf q_h}+{\bf Q}\right|^2}
{\left({\hbar^2(n^2-m^2)\pi^2\over 2a^2m_h^2}+{\hbar^2({\bf q_h}
+{\bf Q})^2\over 2m_h}-{\hbar^2 q_h^2\over2m_h}+\hbar\omega_{ LO}
\right)^2+\Gamma^2}J_{n,m}(Q)\frac{p_m}{\sum_ip_i},
\label{eq:92b}
\end{eqnarray}
where
\begin{displaymath}
p_m=e^{-k_{hm}^2\over q_T^2},\;\; q_h={\sqrt\pi\over2}
\sqrt{2m_hT\over \hbar^2}
\end{displaymath}
Expressions (\ref{eq:92a}) and (\ref{eq:92b}) can be analyzed
easily when the temperature is much lower than the optical phonon
energy. In this case the thermal momenta of holes $q_h$ may
be neglected as compared with the phonon momentum $Q$ which
is approximately equal to the momentum of virtual hole.
It is readily seen that the probability of Auger
transition as a function of $Q$ for phonon emission has
two extrema. The first of these corresponds to the
minimum of the denominator in ($\ref{eq:92a}$) and is related
to a resonant Auger transition. Note that for an Auger
transition with phonon absorption no extremum of this
kind is observed and there is no resonant process. The
second extremum corresponds to the maximum of the squared matrix
element and, as a rule, is related to a virtual Auger
transition. For sufficiently wide quantum wells the
matrix element of the electron-electron interaction as a
function of the heavy hole momentum has a form close to
the $\delta$-function. In this case the probability of the
transition is the highest near the threshold momentum,
and the process of scattering by phonons is virtual. With
decreasing quantum well width, the $\delta$-function broadens for
the quasi-threshold matrix element, and, in addition, the
role of the thresholdless matrix element, only slightly
depending on $Q,$ becomes more significant. This enhances
the resonant Auger transition and weakens the virtual
process. For narrow quantum wells the matrix element of
Coulomb electron-electron interaction depends on $Q$ only
slightly, and, therefore, the resonant process is
predominant. It can readily be shown that in this case
the following estimation is valid for the AR coefficient
of the phonon-assisted Auger process:
\begin{equation}
C_{ph}\approx\frac{\omega_{LO}e^2m_ha}
{8\tilde\kappa\hbar\Gamma}J_{1,1}(Q_0)
\frac{2\pi}{\hbar}{3k(E_g)\over 4E_g}\left|M_{ee}(Q_0)\right|^2,
\end{equation}
where $\textstyle Q_0=\sqrt{2m_h\omega_{LO}\over\hbar}$.
Hence immediately follows that the phonon-assisted to direct
AR coefficient ratio has the form
\begin{equation}
{C_{ph}\over C}\approx\frac{\Gamma_{hp}}{\Gamma}\,\frac{\left(M_{ee}(Q_O)
\right)^2}{\left(M_{ee}(q_T)\right)^2},
\end{equation}
where $C=C_1+C_2$ is the Auger coefficient for the direct process,
$\Gamma_{hp}={\hbar\over\tau_{hp}},\;\tau_{hp}$ is the time of a
hole scattering by an optical phonon; $q_T$ is the thermal momentum of holes.
It can be seen that the phonon-assisted Auger process may
dominate over the direct one only in the case when the
the values of $\Gamma_{ph}$ and $\Gamma$ are close to each other or
at extremely low temperature when the ratio of the matrix elements
taken at the momenta $Q_O$ and $q_T$ is large.
Note that at high nonequilibrium carrier densities,
when Auger recombination becomes at all significant, the
hole-hole scattering turns out to be, as a rule, a much
more effective mechanism of relaxation than the hole-phonon scattering.
This results in a small ratio $\Gamma_{ph}/\Gamma$.
Therefore, the direct Auger recombination dominates over
the phonon-assisted process down to very low
temperatures. Figure 7 shows the coefficients of a phonon-assisted
Auger transition ($C_{ph1}$ and $(C_ph2)$) corresponding to
the thresholdless and quasi-threshold matrix elements
as a function of temperature for different quantum
well widths. As a halfwidth G is taken a characteristic value of
20 meV.

\begin{center}
\section*{\normalsize{\bf VI Discussion}}
\end{center}

Our analysis has shown that for the CHCC and CHHS processes
in semiconductor structures with quantum wells there
exist three AR mechanisms: threshold, quasi-threshold,
and thresholdless. The coefficients of these processes
$C_1$, $C_2$, and $C_3$ depend in different ways both on
temperature and quantum well parameters: the heights of
heterobarriers for electrons and holes ($V_c$ and $V_v$) and
the well width (see Figs. 3 -- 5). In the limit
$a\longrightarrow\infty$ the sum of the
quasi-threshold and the threshold AR coefficients multiplied
by squared quantum well width, $C_2\,a^2+C_3\,a^2$, approaches the
bulk AR coefficient $C^{3D}$, with the product $C_1\,a^2$ approaching
zero (Fig. 8). For sufficiently narrow quantum wells the
2D AR coefficient multiplied by $a^2$ exceeds the 3D value,
owing to the predominance of the thresholdless and quasi-threshold
AR processes. Thus, Auger recombination in quantum wells is
enhanced as compared with that in a homogeneous semiconductor.
This enhancement is more pronounced at low temperature. Under
these conditions the 3D AR coefficient $C^{3D}$ is small because
of the presence of an exponentially small multiplier
(see eq. (\ref{eq:84})). Note that the entire analysis of the AR
coefficients ($C_1$, $C_2$, $C_3$) as functions of temperature and
quantum well parameters is qualitatively applicable to
the same extent to both the CHCC and CHHS Auger
processes. However, since no model structures with
quantum wells have been specified, we illustrated these
relations by the example of the CHCC process.

Note that the AR in quantum wells may be suppressed
substantially if the following conditions are met
$(V_c, V_v)>E_g$ and $E_2-E_1>E_g$ ($E_1$ and  $E_2$ are the energies of
the first and second carrier size quantization levels )
~\cite{25a}; i.e., in the case when the energy of an excited
particle is insufficient for a transition into the
continuous spectrum or to a next size quantization level.
For these conditions to be fulfilled, a structure is to be
created with deep and narrow quantum wells for both
electrons and holes. The presently existing technologies
can produce structures of this kind based on InAs/AlSb ~\cite{25b}
and InAs/GaSb/AlSb~\cite{25c}. In these deep quantum wells only
the threshold AR mechanism, corresponding to the
coefficient $C_3$, is operative. This coefficient may be
smaller by several orders of magnitude than the Auger
coefficients for the thresholdless and quasi-threshold
processes ($C_1$, $C_2$) in shallow quantum wells $((V_c, V_v)<E_g)$.

It is also shown that the phonon-assisted AR process
undergoes significant changes. Similarly to the direct AR
there exist three different phonon mechanisms
($C^3_{ph},\,C^2_{ph},C^1_{ph}$) corresponding to the
threshold, quasi-threshold, and thresholdless
matrix elements of electron-electron interaction. The
first process is quite similar to its 3D counterpart. However,
for narrow quantum wells this process is much weaker than
the thresholdless and threshold Auger processes. It is
this process with the participation of phonons that is
considered in the literature to be principal AR process in
quantum wells~\cite{25,26}. AR phonon-assisted processes
with the quasi-threshold and the thresholdless matrix elements
of electron-electron interaction may be resonant. At low
temperature they can compete with direct AR processes. However,
owing to the lack of any strong temperature dependence in the latter,
such a competition is possible at much lower temperatures
than in the 3D case (see Fig. 7). Thus in narrow quantum
wells the direct (thresholdless) Auger recombination
dominates over the phonon-assisted process in a wider
interval of temperatures than in the 3D case. With
increasing quantum well width, the resonant scattering by
phonons becomes weaker, and we pass to the conventional
3D conditions.

It should be emphasized once again that at high
densities of nonequilibrium carriers in a
homogeneous semiconductor the phonon-assisted AR process
may be less intensive than the Auger recombination, with
the subsequent hole-hole scattering eliminating the
threshold ~\cite{22}.

\begin{center}
\section*{\normalsize{\bf VII Conclusions}}
\end{center}

The following principal results were obtained in the work
\begin{enumerate}
\item It is shown that three main mechanisms of Auger
recombination are operative simultaneously in quantum
wells:
\begin{enumerate}
\item threshold
\item quasi-threshold
\item thresholdless
\end{enumerate}
The first two pass, in the limit of infinitely wide
quantum well, into 3D Auger recombination, and the
coefficient of the thresholdless Auger process tends to zero.
\item It is demonstrated that in the case of narrow quantum
wells the Auger coefficients of the quasi-threshold and
thresholdless processes show weak power-law dependence on
temperature. Their values much exceed the 3D
coefficient related to the squared quantum well width. At
the same time the coefficient of 2D threshold Auger
recombination (a) has a higher threshold energy than in
the 3D case ($E_{th}^{2D}>E_{th}^{3D}$), with the
corresponding AR coefficient smaller than the 3D expression
divided by the squared quantum well width.
\item It is shown that the quasi-threshold Auger process
predominates in wide enough quantum wells. The threshold
energy of this process is an increasing function of temperature
(see Fig. 2). The critical quantum well width at which the threshold
energy for the quasi-threshold Auger process becomes equal to the
bulk value, is found. The critical width $a_c$ strongly (exponentially)
depends on temperature. At room temperatures its value may be up
to several thousand Angstroms for semiconductors with
$E_g\approx\,1\,eV$.
\item In narrow quantum wells the phonon-assisted AR
process is resonant. At high carrier densities it is
less intensive than the direct thresholdless Auger
process down to very low temperature.
\end{enumerate}
\begin{center}
\section*{\normalsize{\bf Acknowledgments}}
\end{center}

The authors would like to thank R.A. Suris and V.I. Perel for
initiating this work. The work was supported in part
by the Russian Foundation for Basic Research (grant nos. 96-02-17952
and 97-02-18151), INTAS (grant no. 94-1172), and Russian State
Program "Physics of Solid State Nanostructures."
\vspace{2cm}

\appendix

\section{ Wave functions of carries in a rectangular quantum well }

\subsection{Holes}

Selecting the coordinate system so that the longitudinal component of the
wave vector coincide with the $y$ axis and performing a Fourier transform in
this plane we obtain the following expressions for the wave functions of
carriers.

Heavy holes:
\begin{eqnarray}
\bbox{\psi}_h(q,x)&=&H_1\left( \begin{array}{c}
q\cos{k_hx}\;\xi\\
-ik_h\sin{k_hx}\;\xi
\label{eq:32}\\
-k_h\sin{k_hx}\;\xi+q\cos{k_hx}\;\eta\end{array}
\right) +\nonumber\\
&&\\
&+&H_2\left( \begin{array}{c}
q\sin{k_hx}\;\eta\\
ik_h\cos{k_hx}\;\eta\\
-q\sin{k_hx}\;\xi-k_h\cos{k_hx}\;\eta\end{array}
\right) .\nonumber
\end{eqnarray}
Where
$\xi=1/\sqrt2 \left( \begin{array}{r}
1\vspace{-3 mm}\\-1 \end{array} \right) ,\hskip 1em \eta=1/\sqrt2
\left( \begin{array}{c} 1 \vspace{-3 mm}\\1 \end{array} \right)$, $H_1$ and
$H_2$ are the normalizing constants.

Light holes:
\begin{eqnarray}
\bbox{\psi}_l(q,x)&=&L_1\left( \begin{array}{c}
k_l\sin{k_lx}\;\eta-\lambda_lq\cos{k_lx}\;\xi\\
-iq\cos{k_lx}\;\eta+i\lambda_l k_l\sin{k_lx}\;\xi\\
-\lambda_l k_l\sin{k_lx}\;\xi+\lambda_l q\cos{k_lx}\;\eta\end{array}
\right)+\nonumber\\
\label{eq:33}
&& \\
&+&L_2\left( \begin{array}{c}
-k_l\cos{k_lx}\;\xi-\lambda_l q\sin{k_lx}\;\eta\\
-i\lambda_l k_l\cos{k_lx}\;\eta-iq\sin{k_lx}\;\xi\\
-\lambda_l q\sin{k_lx}\;\xi-\lambda_l k_l\cos{k_lx}\;\eta\end{array}
\right) .\nonumber
\end{eqnarray}
\begin{equation}
\psi_{sl}=\frac{i\hbar\gamma(k_l^2+q^2)}{E_g+\delta-E}[L_1\cos{k_lx}\;\eta+
L_2\sin{k_lx}\;\xi ].
\label{eq:34}
\end{equation}
\begin{displaymath}
\lambda_l={\delta\over E+2\delta-\hbar^2k_l^2/2m_h},\;\;
\end{displaymath}

The wave functions of SO holes are similar to those of light holes.

A change to functions of another symmetry in the above expressions can be
performed by the formal substitution $\xi\leftrightarrow \eta$  for
$|s\rangle$-, $|x\rangle$-, and $|y\rangle$- components and
$\xi\leftrightarrow -\eta$ for $|z\rangle$-components. The wave functions
of carriers in the barrier region may be obatined similarly to
(\ref{eq:32}-\ref{eq:34}).

If wave functions of two or more particles are considered together,
then in general their
$z$-components of quasi-momentum cannot become zero simultaneously. A change
to a function with arbitrary direction of quasi-momentum can be performed
using the rotation matrix:
\begin{equation}
D_{\varphi}=R_{\varphi}\otimes S_{\varphi},
\label{eq:38}
\end{equation}
where $R_{\varphi}$ acts on the coordinate components of the wave function,
and $S_{\varphi}$ on the spinor components. The Euler angles for a rotation
in the plane $yz$ by an angle  $\varphi$ are
\begin{displaymath}
\Phi=-\pi/2\;\;\; ,\Theta=\varphi\;\;\; ,\Psi=\pi/2.
\end{displaymath}
Thus:
\begin{equation}
R_{\phi}=\left[
\begin{array}{cccc}
\;\;1\;\;\; & 0 & 0 & 0\\
0 & 1 & 0 & 0\\
0 & 0 & \cos\varphi & \sin\varphi\\
0 & 0 & -\sin\varphi & \cos\varphi
\end{array}
\right]\;\;\; ,
S_{\varphi}=\left[
\begin{array}{cc}
\cos{\varphi/2} & -i\sin{\varphi/2}\\
i\sin{\varphi/2} & \cos{\varphi/2}
\end{array}
\right] .
\end{equation}
If the vector ${\bf q}$ has components $q\,(0,\cos{\varphi},\sin{\varphi})$
in the coordinate system $x,y,z$, then the wave function can be written in
the form:
\begin{equation}
\psi_{{\bf q}}\equiv\psi_{\varphi}=D_{-\varphi}\psi_0\; .
\label{eq:40}
\end{equation}
The previously found wave function has a zero subscript.
The wave function of heavy holes, found using (\ref{eq:40}),
is written below, as it will be used later.
\begin{eqnarray}
\bbox{\psi}_h(q,x,\phi)&=&H_1\left[\begin{array}{c}
q\cos{k_h x}e^{-i\phi}\,\xi\\
-ik_h\sin{k_h x}\,\xi-q\cos{k_h x}\sin{\phi}\,\eta\\
-k_h\sin{k_h x}\,\xi+q\cos{k_h x}\cos{\phi}\,\eta
\end{array}\right]+\nonumber\\
\label{eq:32a}\\
&+&H_2\left[\begin{array}{c}
q\sin{k_h x}e^{i\phi}\,\eta\\
ik_h\cos{k_h x}\,\eta+q\sin{k_h x}\sin{\phi}\,\xi\\
-k_h\cos{k_h x}\,\eta-q\sin{k_h x}\cos{\phi}\,\xi
\end{array}\right]\nonumber
\end{eqnarray}

The boundary
conditions for hole wave functions can be derived by integrating Kane's
equations across the heteroboundary (see Section II, paragraph 1.C).
We consider generalized Luttinger
parameters $\tilde\gamma_1$ and $\tilde\gamma_2$ also to be continuous
for the sake of simplicity.
Taking into account that
$m_l^{-1}\approx{2\gamma^2\over E_g+\delta-E}\gg m_h^{-1}$ we
obtain continuity conditions at the heteroboundary for the following
quantities:
\begin{eqnarray}
&&1)\;\bbox{\psi}\; ,\nonumber\\
&&2)\;\frac{\partial}{\partial{x}}\bbox{\psi}_\bot\;,
\label{eq:42}\\
&& 3)\;\frac{1}{E_g+\delta-E}{\rm div}\bbox{\psi}\; .\nonumber
\end{eqnarray}

Generally speaking the wave functions of holes in a quantum well are
superposition of three subbands of the valence band: of heavy, light,
and SO holes. However, the last subband strongly, exponentially,
decays away from the heteroboundary with a exponent
$\kappa_{so}\approx\sqrt{{m_h\Delta\over\hbar^2 }\, {4\over 3}}$. As a
consequence, this branch mainly affects the derivative of the wave function
near the heteroboundary, and its influence on the wave function itself is
negligible. It should be emphasized that such an approximation is not
equivalent to using the $4\times 4$ Hamiltonian from the very beginning. We
shall seek the wave function as a superposition of the heavy- and light-hole
subbands. Near the upper edge of the valence band the parameter
$|\lambda_{so}|\approx {m_h\over m_l}\gg 1$. This means that only the first
and the third of the boundary conditions (\ref{eq:42}) are applicable.
In this approximation, light and heavy holes do not mix with each other
and have different spectra.
The heavy hole spectrum coincides with the quantum mechanical spectrum of a
particle in a rectangular quantum well. For states with even and odd
 $|x>$ -components of the wave function of heavy holes the dispersion
equations are following:
\begin{eqnarray}
\tan{k_ha/2}&=&{\kappa_h\over k_h}\,-\,\mbox{for even states}\nonumber\\
\cot{k_ha/2}&=&-{k_h\over\kappa_h}\,-\,\mbox{for odd states}
\end{eqnarray}
For light holes the states
with different parities cannot be separated, and the dispersion equation
becomes more cumbersome:
\begin{eqnarray}
&&\left[\frac{E_g+\delta+V_c-E}{E_g+\delta-E}\frac{k_l^2+q^2}{\kappa_l^2-q^2}
\kappa_l\cot{k_la/2}+k_l\frac{2\lambda_l-1}{2\tilde\lambda_l-1}\right]\times
\nonumber\\
\times&&\left[\frac{E_g+\tilde\delta+V_c-E}{E_g+\delta-E}\frac{k_l^2+q^2}
{\kappa_l^2-q^2}
\kappa_l\tan{k_la/2}-k_l\frac{2\lambda_l-1}{2\tilde\lambda_l-1}\right]=
\\
=&&q^2\left[\frac{2\lambda_l-1}{2\tilde\lambda_l-1}+\frac{E_g+\tilde\delta+
V_c-E}{E_g+\delta-E}
\frac{k_l^2+q^2}{\kappa_l^2-q^2}\right]^2.\nonumber
\end{eqnarray}
Here $\kappa_l$ and $\kappa_h$ denote the moduli of x quasi-momentum
components of light and heavy holes in the barrier region respectively,
\begin{displaymath}
\widetilde{\lambda}_i=
{\tilde{\delta}\over U_v+E+2\tilde{\delta}+\hbar^2\kappa_l^2/2m_h},\;\;
\tilde{\delta}={\tilde{\Delta_{so}}\over 3}.
\end{displaymath}

Note that at $q=0$ the light hole states are also split into states with
different parities. The constants $H_i$ and $L_i$ in (\ref{eq:32}),
(\ref{eq:33}) are determined by normalization conditions. In particular:
\begin{displaymath}
H_i=\frac{1}{\sqrt{q^2+k_h^2}}\frac{1}{\sqrt{a+\frac{1}{\kappa_h}
\frac{q^2}{q^2+k_h^2}}}.
\end{displaymath}

The opposite is the case for SO holes. The components of the wave functions
of light and heavy holes oscillate rapidly, and the contribution from them
to overlap integrals is negligibly small. Similarly, it is easy to verify
that for the $so$-component $\psi_x$ and ${\rm div}{\bf\psi}/(E_g+\delta-E)$
are to be considered continuous. The type of wave functions of SO
holes is similar to that for the light holes (\ref{eq:33}),(\ref{eq:34}).
Strictly speaking, with the condition $E_g-\Delta>U_v$ fulfilled the spectrum
of spin-split holes is continuous. However, when the rapidly oscillating
contributions from the subbands of light and heavy holes are neglected,
the spectrum may be both continuous and discrete. In the general case, near
such a quasi-discrete level there exists a peak of density of states with
small hole momenta in the direction perpendicular to the heteroboundary.
The spectrum of holes of this kind is similar to that of light holes. The
corresponding dispersion equation for SO holes has the form
\begin{eqnarray}
&&\left[\frac{E_g+\tilde\delta+V_c-E}{E_g+\delta-E}\frac{k_{so}^2+q^2}
{\kappa_{so}^2-q^2}
\kappa_{so}\cot{k_{so}a/2}+k_{so}\right]\times\nonumber\\
\times&&\left[\frac{E_g+\tilde\delta+V_c-E}{E_g+\delta-E}\frac{k_{so}^2+q^2}
{\kappa_{so}^2-q^2}
\kappa_{so}\tan{k_{so}a/2}-k_{so}\right]=
\\
=&&q^2\left[\frac{\lambda_{so}}{\tilde\lambda_{so}}+
\frac{E_g+\tilde\delta+V_c-E}{E_g+\delta-E}
\frac{k_{so}^2+q^2}{\kappa_{so}^2-q^2}\right]^2.\nonumber
\end{eqnarray}

\subsection{Electrons}

Electrons obey the same symmetry rules as holes. Their wave functions
are similar to those of light holes and can be written as:
\begin{eqnarray}
&&\mbox{at}\;|x|<a/2\; , \nonumber \\
&&\psi_{sc}=A_1\cos{k_cx}\,\eta+A_2\sin{k_cx}\,\xi ,\nonumber\\
&&\bbox{\psi_c}=\frac{i\hbar\gamma}{Z}A_1\left( \begin{array}{c}
k_c\sin{k_cx}\,\eta-\lambda_cq\cos{k_cx}\,\xi\\
-iq\cos{k_cx}\,\eta+i\lambda_c k_c\sin{k_cx}\,\xi\\
-\lambda_c k_c\sin{k_cx}\,\xi+\lambda_c q\cos{k_cx}\,\eta\end{array}
\right) +\\ \label{eq:45}
&+&\frac{i\hbar\gamma}{Z}A_2\left(
\begin{array}{c}
-k_c\cos{k_cx}\,\xi-\lambda_c q\sin{k_cx}\,\eta\\
-i\lambda_c k_c\cos{k_cx}\,\eta-iq\sin{k_cx}\,\xi\\
-\lambda_c q\sin{k_cx}\,\xi-\lambda_c k_c\cos{k_cx}\,\eta\end{array}
\right) ,\nonumber\\
\end{eqnarray}
where
\begin{eqnarray}
&&Z=\frac{{\cal E}^2+{\cal E}(2E_g+2\delta)+(E_g+3\delta)E_g}
{{\cal E}+E_g+2\delta}\;
\label{eq:47}\\
&&\lambda_c=\frac{\delta}{{\cal E}+E_g+2\delta}\; .\nonumber
\end{eqnarray}
Here $k_c$ denotes the $x$-component of the quasi-momentum of electrons in a
quantum well, $q$ is the longitudinal quasi-momentum of electrons.
Functions with another
symmetry can be derived by the same procedure as that used for holes. From
the boundary condition follows that $\psi_s$ and $\psi_x$ must be continuous.
This yields the following dispersion equation:
\begin{equation}
\left(k_c\tan{k_ca/2}-\frac{Z}{\tilde{Z}}\kappa_c\right)\left(k_c\cot{k_ca/2}
+\frac{Z}{\tilde{Z}}\kappa_c\right)=-q^2\left(\lambda_c-\tilde{\lambda_c}
\frac{Z}{\tilde{Z}}\right)^2,
\label{eq:49}
\end{equation}
where $\kappa_c$ is the modulus of the $x$ quasi-momentum component of
electrons in the barrier region,
\begin{eqnarray*}
&&\tilde{Z}=\frac{{\cal E}^2+{\cal E}(2E_g+2U_v+2\tilde{\delta})+
(E_g+U_v+3\tilde{\delta})(E_g+U_v)}
{{\cal E}+E_g+U_v+2\tilde{\delta}}\\
&&\tilde{\lambda_c}=\frac{\tilde{\delta}}{{\cal
E}+E_g+U_v+2\tilde{\delta}}\;.
\end{eqnarray*}

The spectrum splits into even and odd states if the longitudinal wave vector
($q$) is small or the expression in parentheses in the right-hand side of the
equation is close to zero. The last condition is commonly fulfilled, since,
as a rule,  $U_v\ll E_g$, which corresponds to semiconductors with about the
same band structure. Note, that in the case of the discontinuous
Kane parameter $\gamma\neq const$, continuity of $\gamma \psi_x$
and $\psi_s$ should be used~\cite{16}.

\section{Coulomb potential in the presence of heteroboundaries}

In a quantum well the Coulomb potential of a charged particle
differs from that in a homogeneous semiconductor, because of the
different dielectric constants in a quantum well and barrier region:
\begin{equation}
\Phi({\bf r}_0,{\bf r})={e\over \kappa_0|{\bf r}-{\bf r}_0|}+
\tilde\Phi({\bf r}_0,{\bf r}),
\end{equation}
where ${\bf r}_0$ is the coordinate of the particle and ${\bf r}$ is the
coordinate of the point where the potential is observed. We consider
only the case, when the particle is inside the quantum well
($|x_0|<a/2$). Using the reflection method (see e.g.~\cite{VIII}) one
can obtain:
\begin{eqnarray}
\tilde\Phi &=&\sum_{n\geq 1}{e\over \kappa_0}
\left({\kappa_0-\tilde\kappa_0\over
\kappa_0+\tilde\kappa_0}\right)^{2n-1}\left( {1\over
\sqrt{(x+x_0-(2n-1)a)^2+\rho^2}}+{1\over
\sqrt{(x+x_0+(2n-1)a)^2+\rho^2}}\right)+\nonumber\\
&+&\sum_{n\geq 1}{e\over \kappa_0}
\left({\kappa_0-\tilde\kappa_0\over
\kappa_0+\tilde\kappa_0}\right)^{2n}
\left({1\over
\sqrt{(x-x_0-2na)^2+\rho^2}}+{1\over
\sqrt{(x-x_0+2na)^2+\rho^2}}\right)\;\;\mbox{at}\;\;|x|<a/2\nonumber\\
\tilde\Phi &=&{e\over \kappa_0\sqrt{(x-x_0)^2+\rho^2}}
{\kappa_0-\tilde\kappa_0\over \kappa_0+\tilde\kappa_0}+
{2e\over \kappa_0+\tilde\kappa_0}
\sum_{n\geq 1} \left({\kappa_0-\tilde\kappa_0\over
\kappa_0+\tilde\kappa_0}\right)^{2n}{1\over
\sqrt{(x-x_0+2na)^2+\rho^2}}+\label{Phi}\\
&+&{2e\over \kappa_0+\tilde\kappa_0}
\sum_{n\geq 1}\left({\kappa_0-\tilde\kappa_0\over
\kappa_0+\tilde\kappa_0}\right)^{2n-1}{1\over
\sqrt{(x+x_0+(2n-1)a)^2+\rho^2}}\;\;\;\;\mbox{at}\;\;x>a/2.\nonumber
\end{eqnarray}
Here $\rho^2=(y-y_0)^2+(z-z_0)^2$, $a$ is a quantum well width.
These potentials are rather cumbersome. However, they may be
simplified if dielectric constants $\kappa_0$ and $\tilde\kappa_0$
are supposed to be close to each other. After taking Fourier
transforming along lateral coordinates ($y$ and $z$) one obtains:
\begin{eqnarray}
\phi(x,q)&\approx& {e\over 2q\kappa_0}
\left(e^{-q|x-x_0|}+2{{\kappa_0-\tilde\kappa_0\over
\kappa_0+\tilde\kappa_0}}{\rm ch}(q(x+x_0))e^{-qa}\right)
\;\;\mbox{at}\;\;|x|<a/2\\
\phi(x,q)&\approx& {e\over q (\kappa_0+\tilde\kappa_0)}\left(
e^{-q(x-x_0)}+{\kappa_0-\tilde\kappa_0\over
\kappa_0+\tilde\kappa_0} e^{-q(x+x_0+a)}\right)\;\;\;
\mbox{at}\;\;x>a/2.\nonumber
\end{eqnarray}
One may see that while potential itself is a continuous function
along the interface, difference between its left and right
derivatives is proportional to $(\kappa_0-\tilde\kappa_0)/
(\kappa_0+\tilde\kappa_0)$.

\newpage
\begin{center}
{\Large Footnotes}
\end{center}
\begin{enumerate}

\item We assume that, as is commonly the case, $(V_c,\, V_v)<E_g$

\item  The reason is that this term has the lowest
threshold energy. By the threshold energy we understand the mean energy of a
heavy hole taking part in an Auger transition

\end{enumerate}

\newpage
\begin{center}
{\Large Figure Captions}
\end{center}

Fig. 1 Auger transition probabilities
$w_1(q)$ and $w_2(q)$, corresponding to the thresholdless and
quasi-threshold matrix elements $M_1$ and $M_2$ as functions of
the longitudinal
momentum of heavy holes at $T=300\,K$ for different quantum well
widths ((a) $a=50\,$\AA, (b) $a=100\,$\AA, (c) $a=200\,$\AA, and (d)
$a=500\,$\AA). The model parameters typical of InGaAsP/InP quantum
wells with $E_g\approx 1\,eV$ were used.\\
\\
Fig.2 Threshold energy of the CHCC process as a function of the
quantum well width for three mechanisms of Auger recombination:
thresholdless ($E_{th}^1$), quasi-threshold ($E_{th}^2$), and threshold
($E_{th}^3$) at $T=300\,K$. The solid curve corresponds to the threshold
energy ($E_{th}^{tot}$) of the total Auger coefficient ($C=C_1+C_2+C_3$).
The horizontal dashed line corresponds to the threshold energy
$E_{th}^{3D}$ for the bulk Auger process.\\
\\
Fig. 3 Auger coefficients $C_1$ and $C_2$ for the thresholdless and
quasi-threshold processes as functions of a quantum well width at different
temperatures ((a) $T=50\,K$, (b) $T=150\,K$, (c) $T=300\,K$, and (d)
$T=400\,K$). The same parameters as in Fig. 1 were used \\
\\
Fig. 4 Coefficient $C_3$ for the threshold Auger process as a function of
the quantum well width at $T=150\,K$ (a) and $T=300\,K$ (b).\\
\\
Fig. 5 Temperature dependence of the total Auger coefficient and the
partial contributions of the thresholdless and quasi-threshold mechanisms
at different quantum well widths. The same parameters were used as in
Fig. 1.\\
\\
Fig. 6 Comparison of the thresholdless Auger coefficient ($C_1$)
as a function of temperature at various Fermi energies of the
holes for two different quantum well thicknesses ((a) $a=50\,$\AA$\;$ and
(b) $a=500\,$\AA). $T_F$ denotes the Fermi energy expressed
in Kelvins. The curve with $T_F=-100\,K$ approximately corresponds to
the Boltzmann statistics.\\
\\
Fig. 7 Phonon-assisted Auger coefficients corresponding to the
thresholdless and quasi-threshold matrix elements as functions of
temperature at different quantum well widths ((a) $a=50\,$\AA$\;$ and
(b) $a=150\,$\AA). \\
\\
Fig. 8 Three-dimensional Auger coefficients $C_1\,a^2$ and $(C_2+C_3)\,a^2$
as functions of a quantum well width at $T=300\,K$.
The same parameters as in Fig. 1 were used.

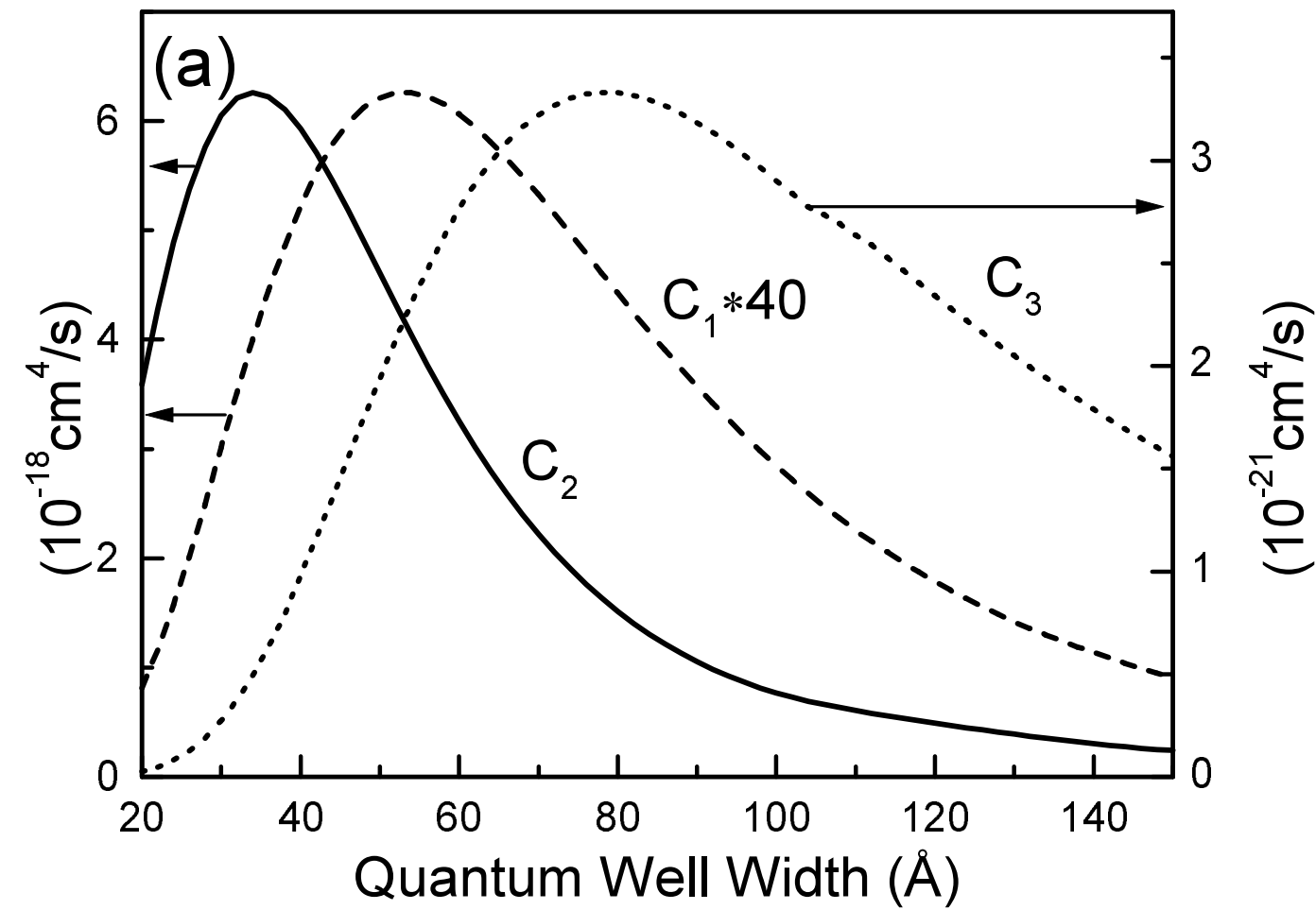
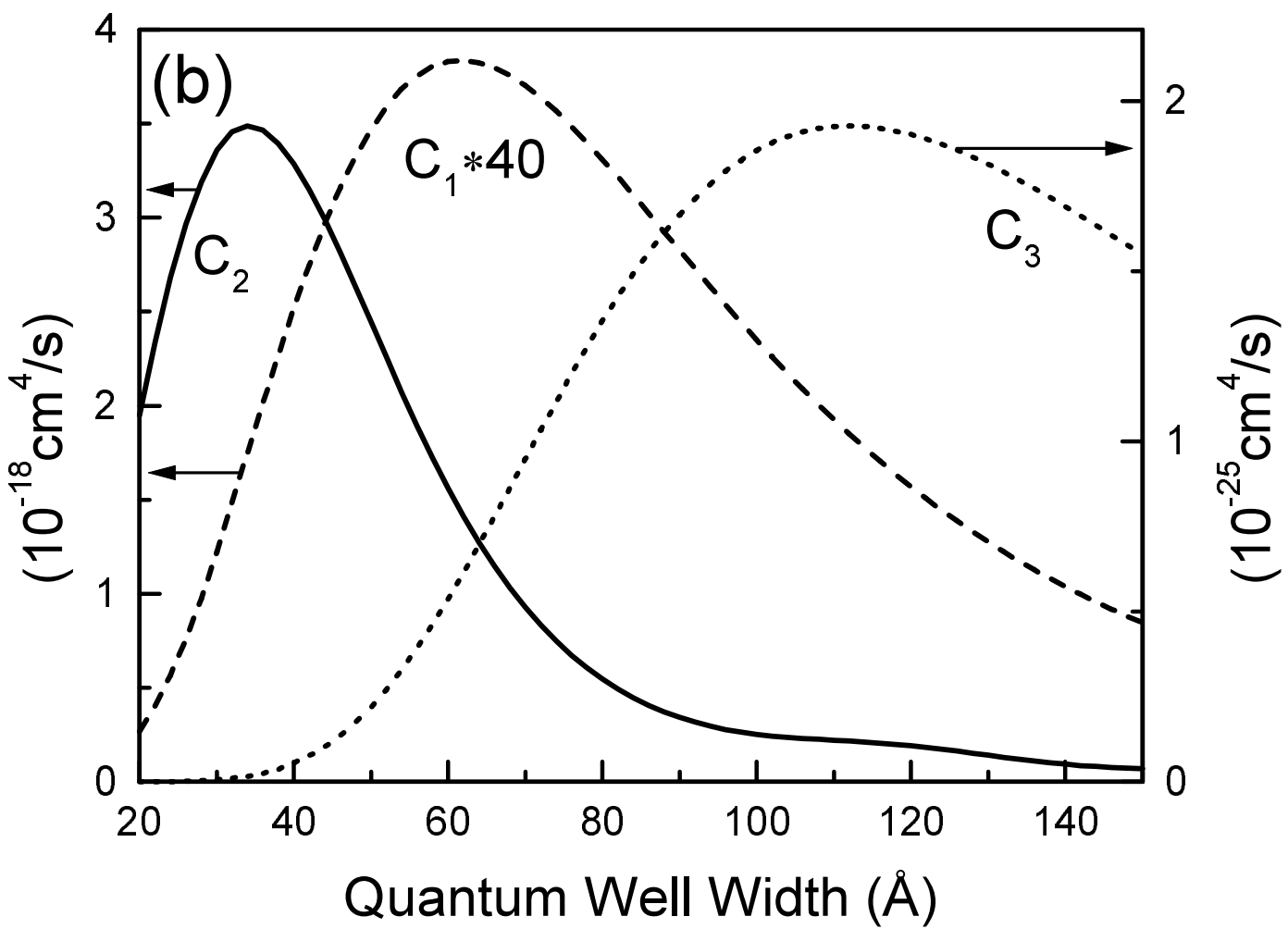
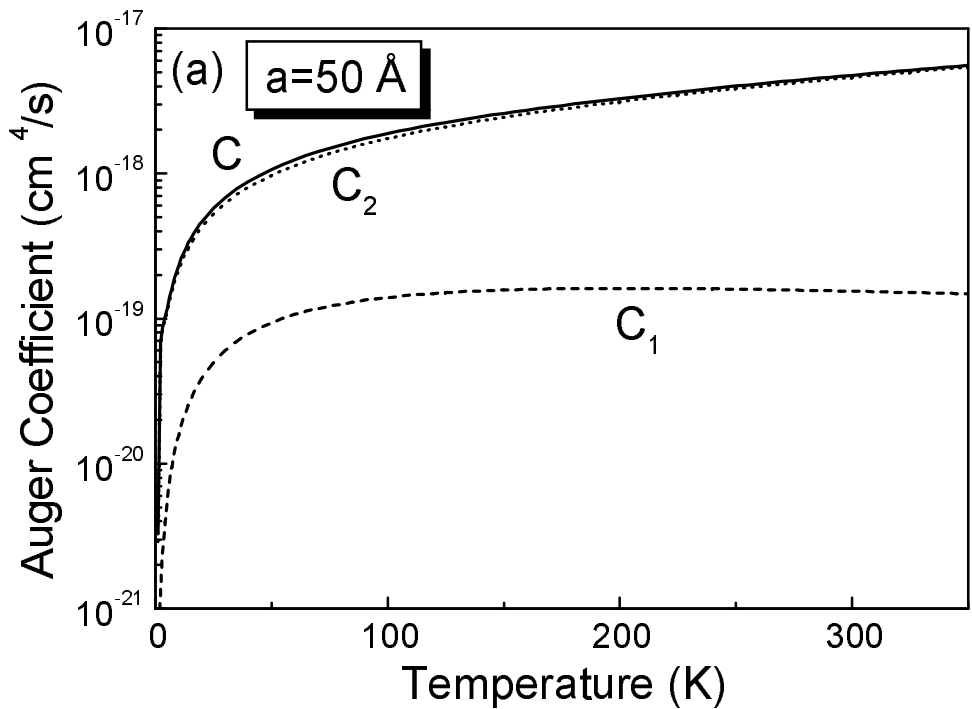
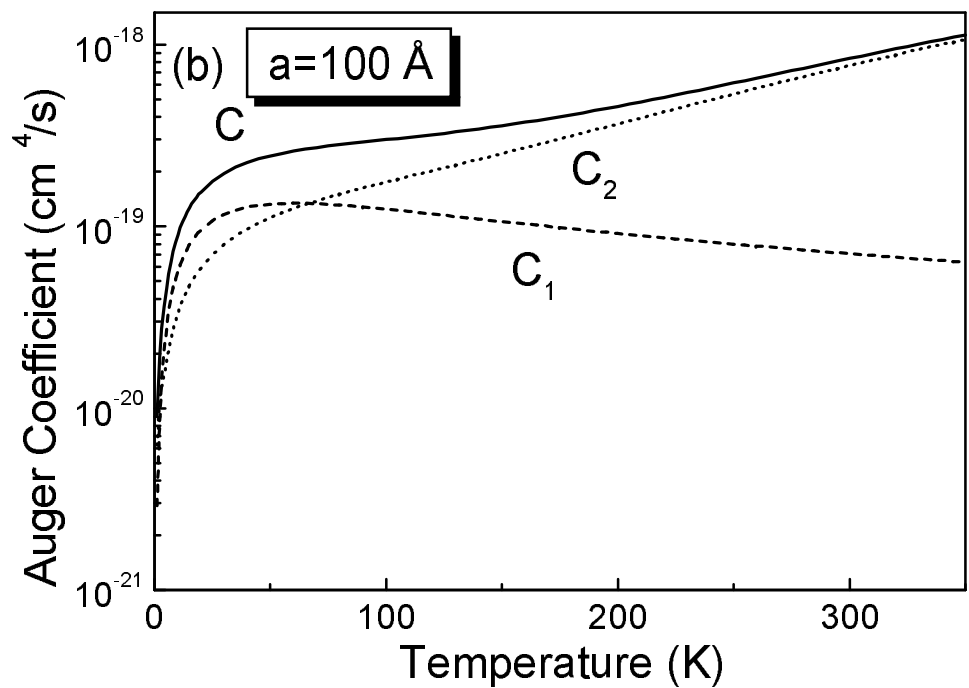
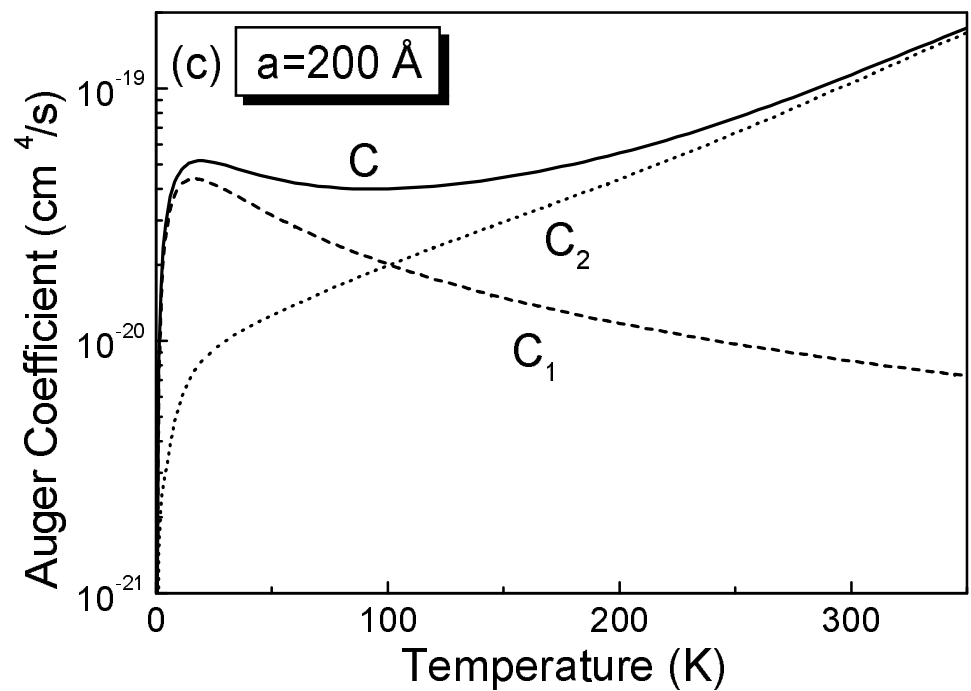
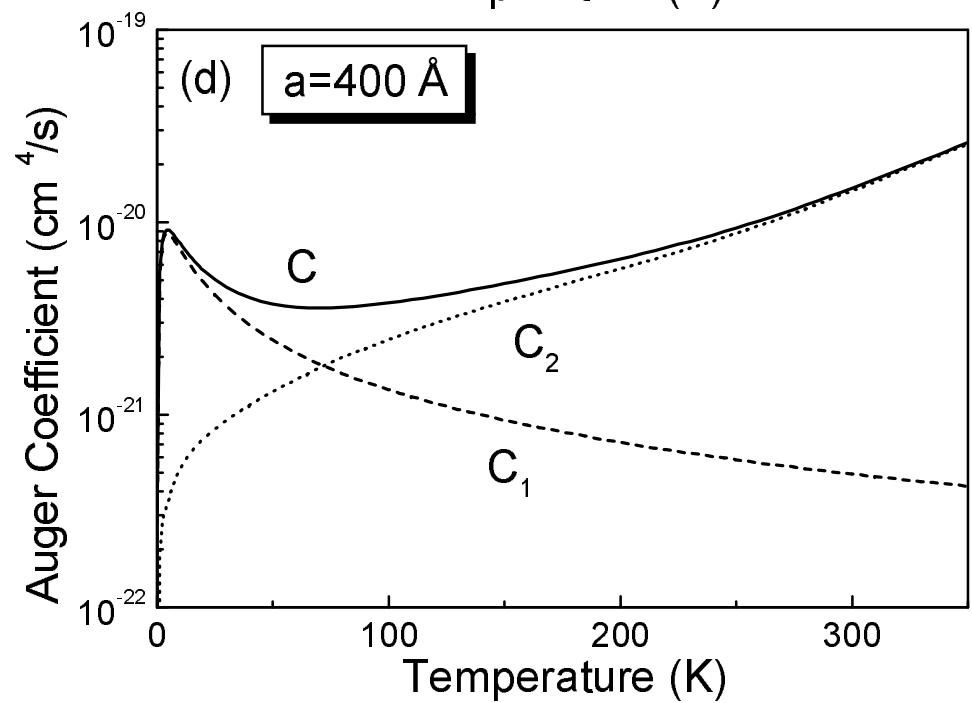
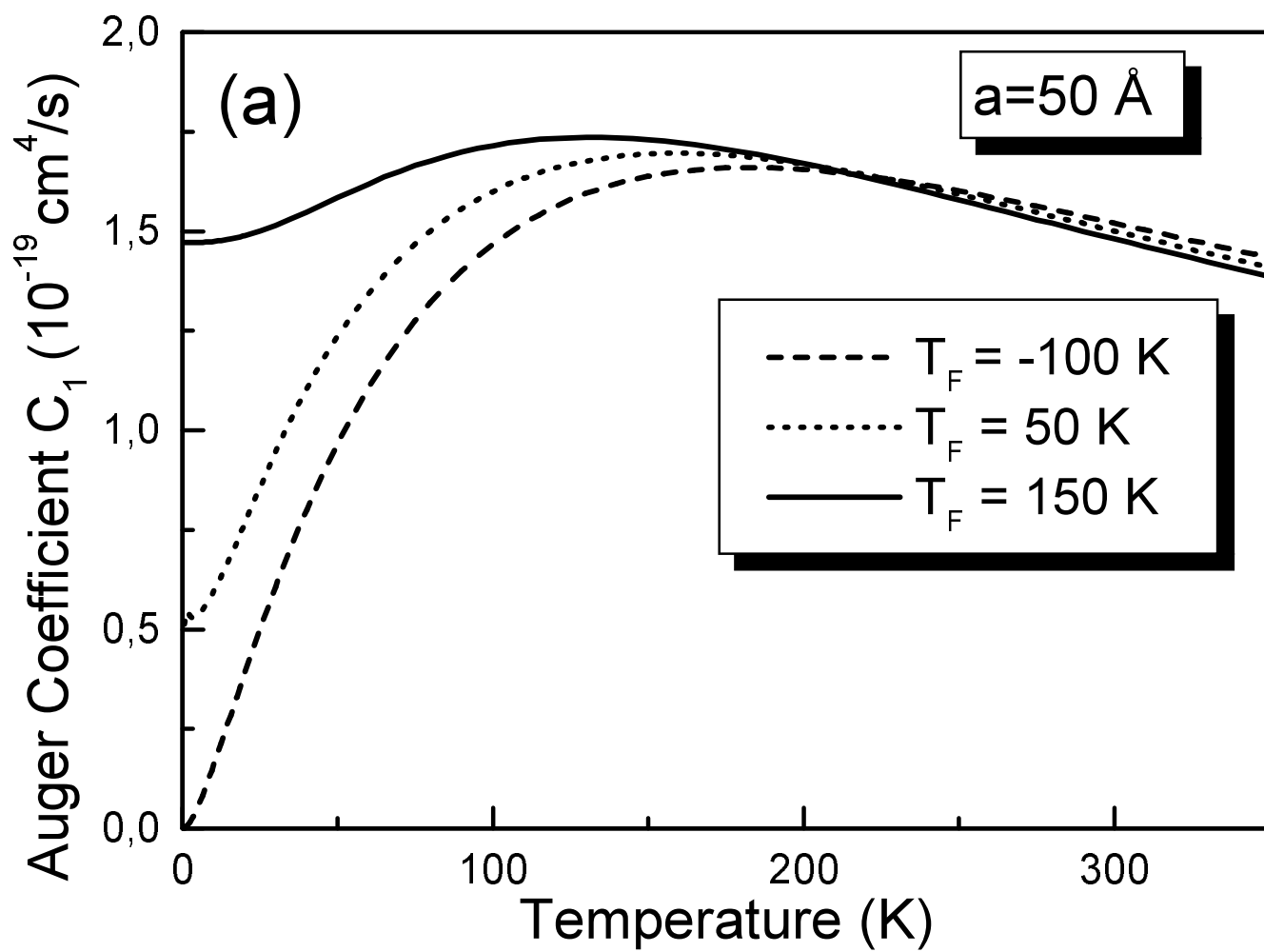
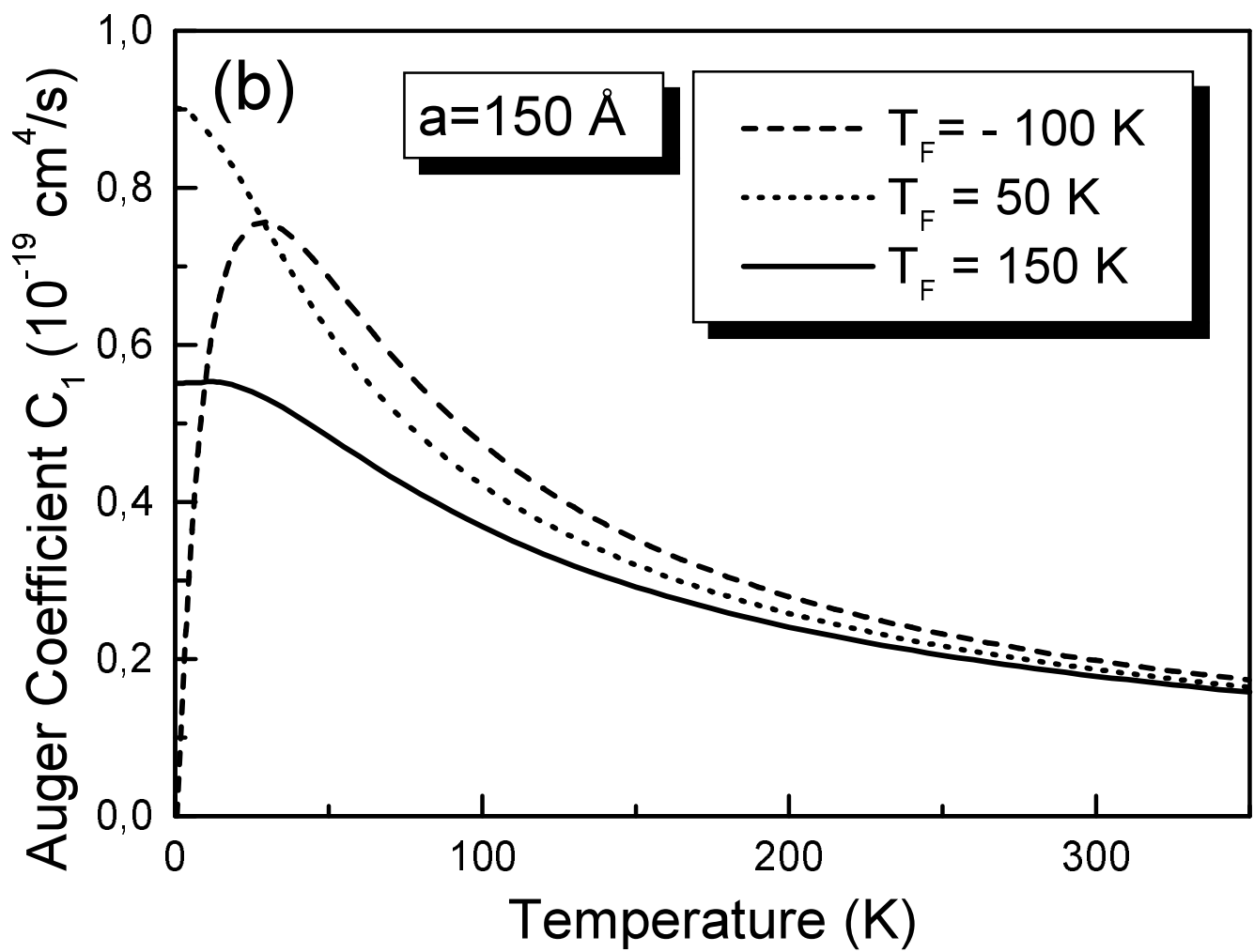
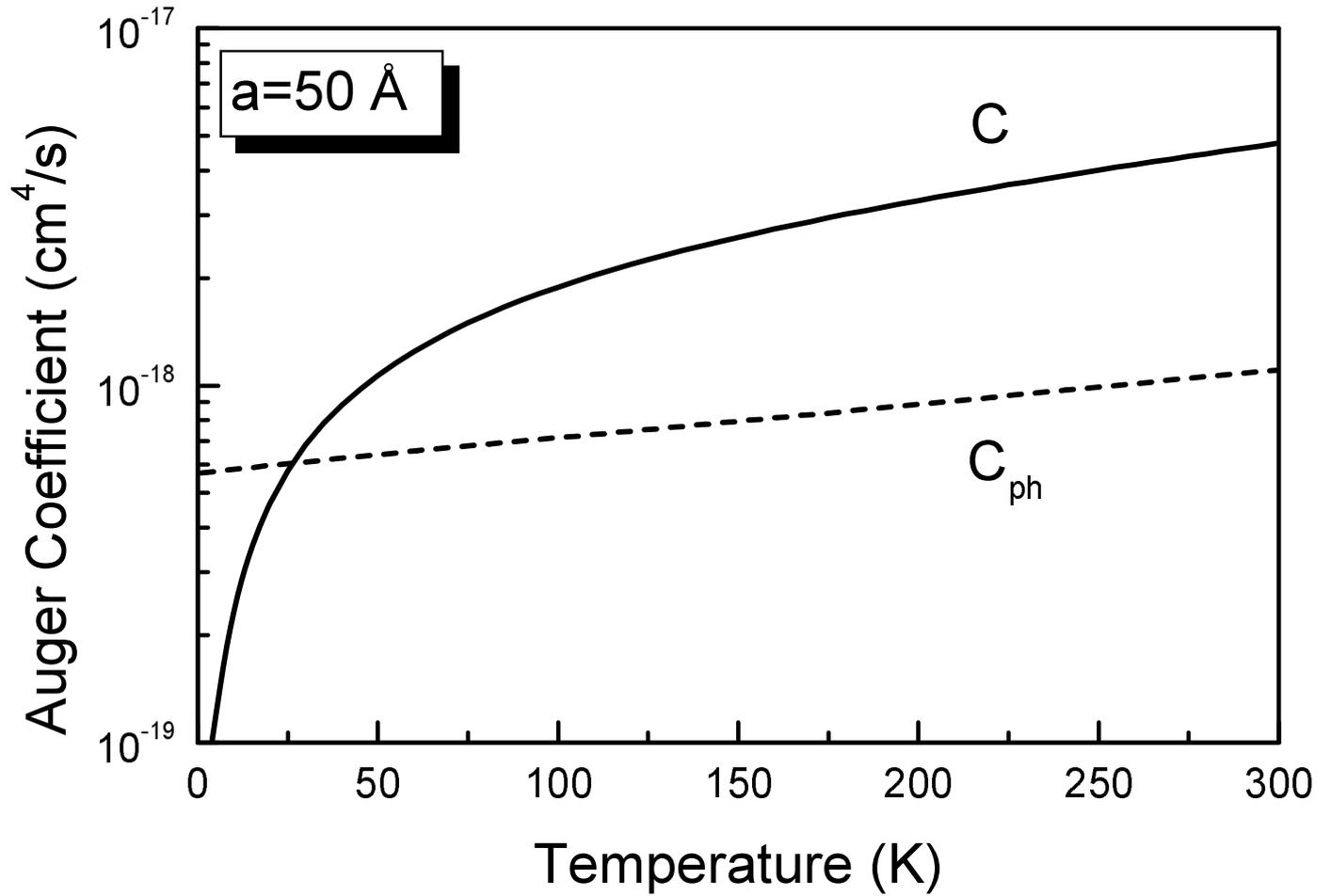
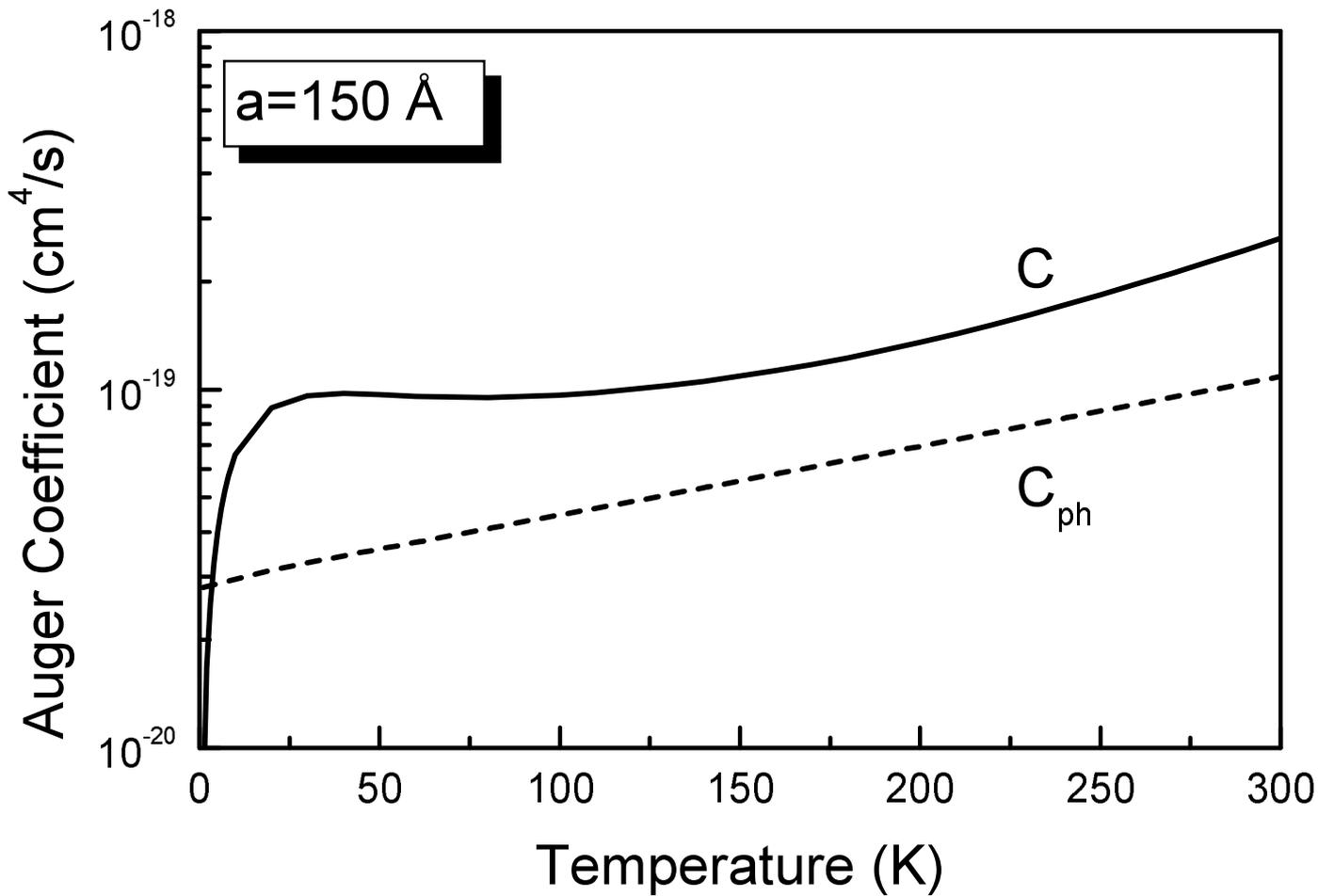
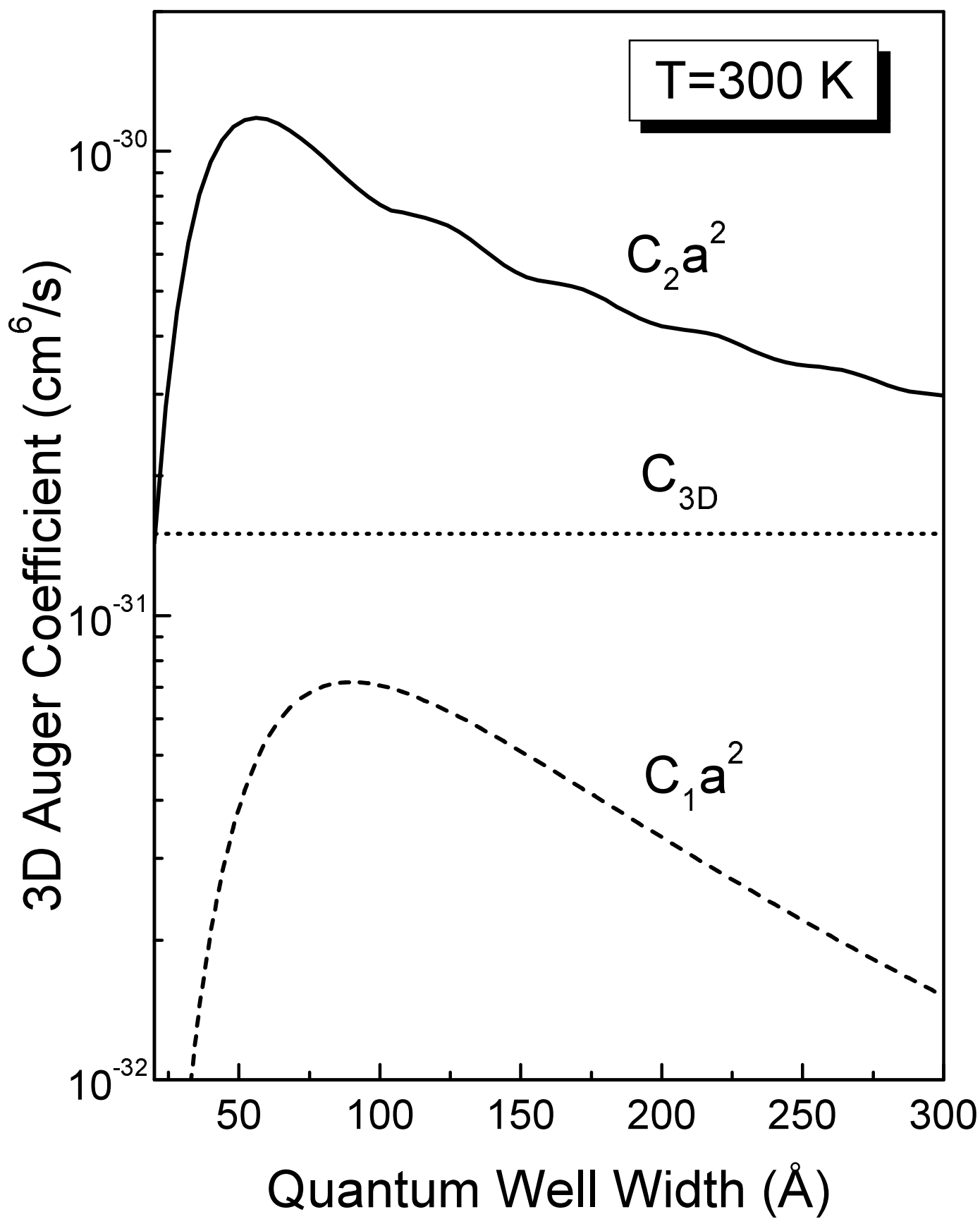
\end{document}